\documentclass[final]{svjour3}                     % onecolumn (standard format)
\smartqed  % flush right qed marks, e.g. at end of proof
\usepackage{graphicx}
\usepackage{amsmath} %for \eqref
\usepackage{color}
\usepackage{cite}
\usepackage{hyperref}

\hypersetup{
pdfborder={0 0 0}, %omits the borders for links
colorlinks=true, % false: boxed links; true: colored links
linkcolor=blue,
citecolor=blue,
breaklinks=true, %for standard latex (dvi->ps->pdf) allow text in links to have a line break
urlcolor=blue
}
\usepackage{cite}
\journalname{Journal of Low Temperature Physics}

\begin{document}

\title{On the Transition from Potential Flow to Turbulence Around a Microsphere Oscillating in Superfluid $^4$He}%\thanks{Grants or other notes

\titlerunning{Oscillating Sphere}        % if too long for running head

\author{M. Niemetz \and R. H\"anninen  \and W. Schoepe} 
        %Second Author %etc.


\institute{M. Niemetz \at OTH Regensburg, Regensburg, Germany \and R. H\"anninen \at Low Temperature Laboratory, Department of Applied Physics, Aalto University, Finland \and
W. Schoepe \at Fakult\"at f\"ur Physik, Universit\"at Regensburg, Regensburg, Germany\\
\email{wilfried.schoepe@ur.de}}

                      %  \\
           %\and
           %S. Author \at
              %second address

\date{}

\maketitle

\begin{abstract}
The flow of superfluid $^4$He around a translationally oscillating sphere, levitating without mechanical support, can either be laminar or turbulent, depending on the velocity amplitude.
Below a critical velocity $v_c$ that scales as $\omega ^{1/2}$, and is temperature independent below 1\,K, the flow is laminar (potential flow).
Below 0.5\,K the linear drag force is caused by ballistic phonon scattering that vanishes as T$^4$ until background damping, measured in the empty cell, becomes dominant for T $<$ 0.1\,K. 
Increasing the velocity amplitude above $v_c$ leads to a transition from potential flow to turbulence, where the large turbulent drag force varies as $(v^2 - v_c^2)$. 
In a small velocity interval $\Delta v / v_c \le 3 \%$ above $v_c$, the flow is unstable below 0.5\,K, switching intermittently between both patterns. 
From time series recorded at constant temperature and driving force, the lifetimes of both phases are analyzed statistically. 
We observe metastable states of potential flow which, after a mean lifetime of 25 minutes, ultimately break down due to vorticity created by natural background radioactivity. 
The lifetimes of the turbulent phases have an exponential distribution, and the mean increases exponentially with $\Delta v^2$. 
%We infer the frequency with which vortex rings are shed from the sphere. 
We investigate the frequency at which the vortex rings are shed from the sphere.
Our results are compared with recent data of other authors on vortex shedding by moving a laser beam through a Bose-Einstein condensate. 
Finally, we show that our observed transition to turbulence belongs to the class of ``supertransient chaos'' where lifetimes of the turbulent states increase faster than exponentially. 
Peculiar results obtained in dilute $^3$He - $^4$He mixtures are presented in the Appendix.

%\PACS{67.25.Dk \and 67.25.Dg \and 47.27.Cn}
\end{abstract}

\section{INTRODUCTION}
\label{intro}
Turbulence in superfluids, often named quantum turbulence, is a very active area of current research \cite{Mak}. 
Quantum mechanical constraints require that only vortices of quantized circulation can exist in an otherwise irrotational flow \cite{Russ}. 
In the limit of very low temperature when almost all thermal excitations are frozen out, turbulence in a pure superfluid is considered to be the most simple paradigm of a difficult subject. 
This applies to superfluid $^4$He at millikelvin  temperatures, to superfluid $^3$He in the microkelvin regime, and to Bose-Einstein condensates (BEC) at nanokelvin temperatures. 
The method to generate vorticity that leads to turbulence in the form of a vortex tangle is to move an object through the superfluid, or to rotate it. 
In superfluid helium using oscillating objects like spheres, wires, grids or tuning forks is the easiest way to study the transition from potential flow to turbulence at some critical velocity amplitude. For recent reviews see, e.g.,   \cite{VSkr,VS}. New techniques are now available to obtain more information of the turbulent state, like improved electronics \cite{Electronics} and visualization by means of Particle Image Velocimetry (PIV) \cite{Zhang,Pao,Guo,Zemma}. 

In case of a BEC the moving object is typically a laser beam that presents an obstacle to the condensate, and thus may lead to vortex shedding above a critical velocity at which the beam is moved repeatedly or back and forth through the condensate. This phenomenon, that was theoretically predicted and observed by several groups, is related to our experiments in the dense Bose liquid of superfluid $^4$He, as will be shown below.

As mentioned above, quantum turbulence can also be investigated by rotating the superfluids. Although this technique will not be discussed in the following, interesting review articles can be found in \cite{Fetter,Bagnato} for BEC, and \cite{Elt} for the helium liquids.

In our present article we review and summarize our earlier experimental and theoretical results from experiments with an oscillating microsphere immersed in superfluid $^4$He at temperatures down to 25 mK. 
The experimental technique (for details, see Appendix A) makes use of superconducting levitation of a ferromagnetic sphere (radius $R$ = 124 $\mu$m, mass $m$ = 27$\mu$g) between superconducting niobium electrodes of a horizontal parallel plate capacitor (spacing $d$ = 1 mm). 
Before cooling the capacitor into the superconducting state (for Nb at 9.2 K), we apply several hundred volts to the bottom electrode charging the sphere to about $q \sim $ 1 pC. 
Vertical oscillations around the equilibrium position of the levitating sphere can be excited by applying an ac voltage $U_{ac}$ at resonance ($\sim$ 120 Hz) in the range from 0.1 mV to several volts, exerting on the sphere a driving force $F = q\,U_{ac}/d$. 
The oscillations induce an ac current $I = q\,v/d$ that is detected by an electrometer. 
Because no mechanical support is needed, the sphere moves at a well-defined velocity in contrast to wires, grids or tuning forks, where the velocity changes from zero at the mechanical support to some maximum value at the center of the wire or at the tip of the fork. Moreover, the simple spherical geometry makes the data transparent and more directly accessible in a quantitative way, in particular the laminar and the turbulent drag forces on the sphere can be identified quantitatively, in contrast to the complicated geometry of the other oscillating objects.\\

Potential flow and the transition to turbulent flow at a sharp critical velocity $v_c$ can be easily identified due to the very different drag forces. 
Most interestingly, in a small interval of velocities $\Delta v = v - v_c$ above $v_c$ where $\Delta v / v_c \le 3 \%$, we observe an instability of the flow pattern switching intermittently between phases of potential flow and turbulent flow. 
An investigation of this phenomenon leads to detailed information on the stabibility of the turbulent state and allows an interesting comparison with vortex shedding in a BEC.\\

This article is organized as follows: in Section 2 the results for stable potential flow and stable turbulent flow will be presented; in Section 3 we review our results for the critical velocity. 
The central part of the article is the intermittent switching phenomenon discussed in Section 4: in Subsection 4.1 the stability of the phases of potential flow is discussed; in Subsection 4.2 the lifetimes of the turbulent phases are analyzed; in Subsection 4.3 vortex shedding from the sphere is compared with experiments of other authors obtained by moving a laser beam through a BEC; in Subsection 4.4 we find from a recent definition of a superfluid Reynolds number that the transition to turbulence in our experiments is a new example of ``supertransient'' chaos. 
After the final Summary, we present experimental details in Appendix A. 
Some peculiar results obtained with dilute $^3$He -$^4$He mixtures are described in Appendix B.
 
\section{STABLE FLOWS}
\label{flow}
\begin{figure}[t]
\centerline{\includegraphics[width=0.90\textwidth]{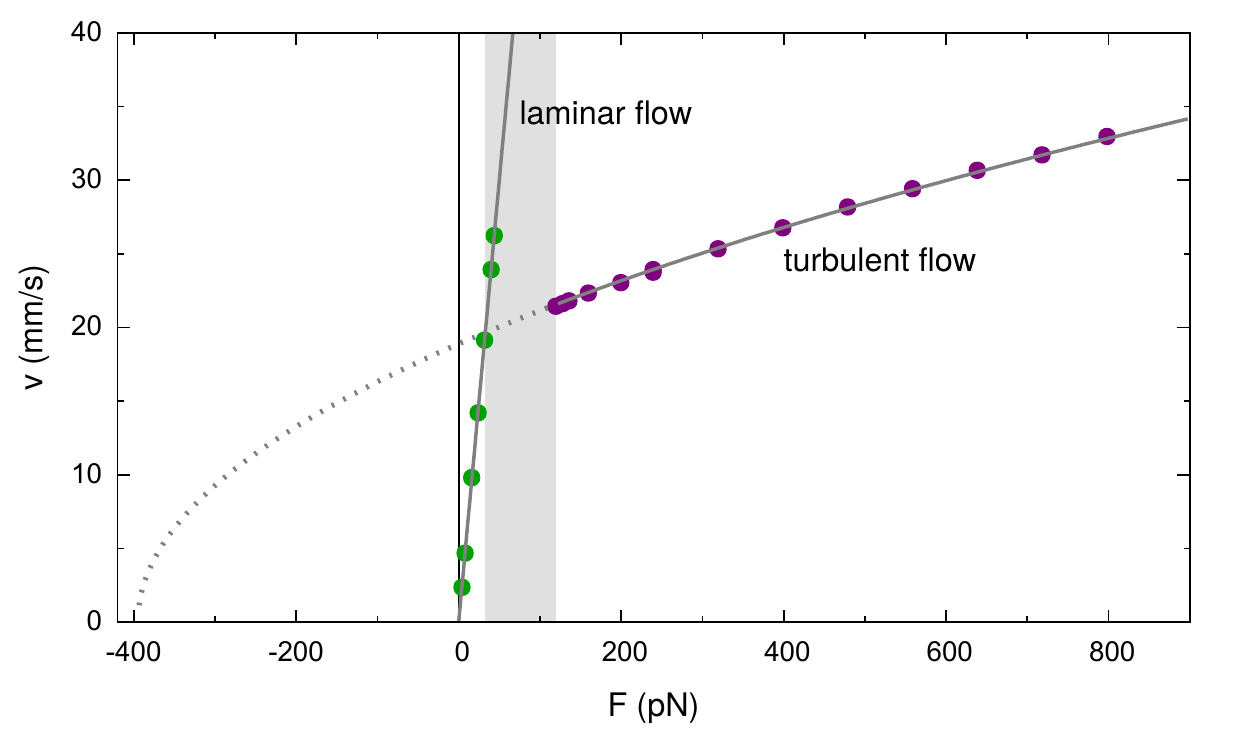}}
\caption{(From \cite{NiemetzSchoepe2004}) Velocity amplitude as a function of the driving force amplitude at 300 mK. 
At small drives the linear increase is the regime of potential flow and the slope is given by ballistic phonon scattering. 
At larger driving forces we observe stable nonlinear turbulent drag, where the solid line is a fit to a quadratic drag force  (linear phonon drag subtracted), see text. 
Note that the apex of the parabola is shifted to the left from the origin. 
The shaded area indicates the unstable regime above a critical velocity where the flow switches intermittently between both patterns. (Color figure online)}
\label{fig:1}
\end{figure}

At a fixed temperature $T$ we measure the velocity amplitude as a function of the amplitude of the driving force $v(F)$ at resonance frequency. 
(Details of the determination of the resonance frequency are described in Appendix A.) 
As an example, the data at $T$ = 300 mK are shown in Fig.\,\ref{fig:1}. 
We identify three different regimes: a linear rise of $v(F)$ at small driving forces up to a critical velocity; a nonlinear increase at larger drives; and the shaded region in between where the flow pattern is unstable, switching between both patterns \cite{NiemetzSchoepe2004}. 
In the following we will discuss these regimes in detail.

\begin{figure}[h]
\centerline{\includegraphics[width=0.80\textwidth]{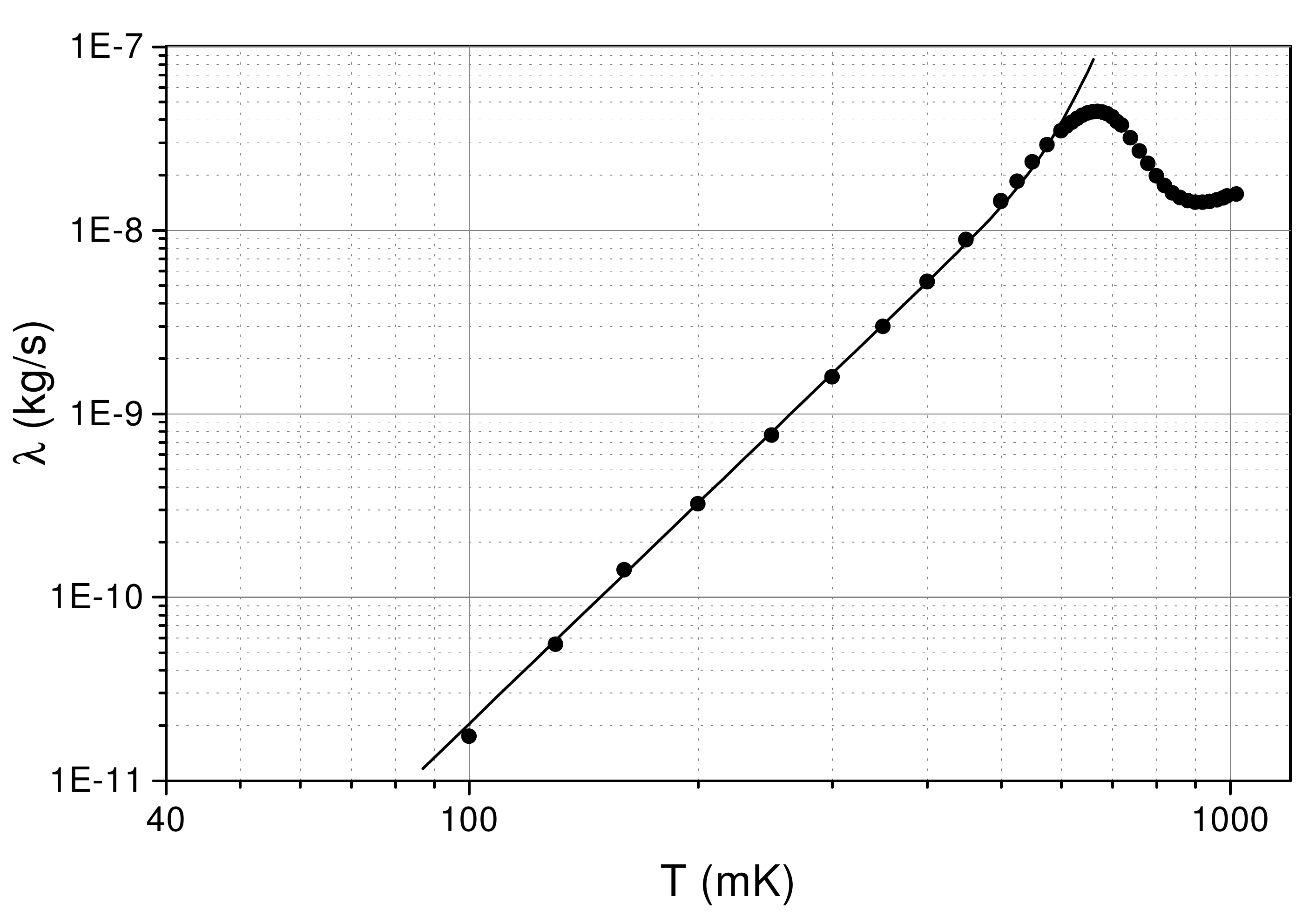}}
\caption{(From \cite{NiemetzSchoepe2004}) Drag coefficient $\lambda(T)$ of the laminar phase. 
The solid line is calculated for ballistic quasiparticle scattering. At the lower temperatures it follows a $T^4$ law due to phonon scattering, see Eq.\,\eqref{Eq:2}. Also note that the drag is larger near 0.7 K than at 0.9 K where the transition to the hydrodynamic regime begins. 
A background damping of $4.4\cdot10^{-11}$\,kg/s measured in vacuum has been subtracted.} 
\label{fig:2}
\end{figure}

\begin{bf}
\subsection{Stable potential flow}
\end{bf}
The linear regime at small velocities indicates potential flow with a small linear drag force
\begin{equation}
F_D = \lambda (T)\,v  \label{Eq:1}
\end{equation}
that, because of the $T^4$ temperature dependence of $\lambda$, we attribute to ballistic phonon scattering, whereby
\begin{equation}
\lambda(T)\,=\,\rho_{\mathsf{ph}}  \cdot c \cdot \pi R^2\,\propto T^4\,,\label{Eq:2}
\end{equation}
where the phonon density $\rho_{\mathsf{ph}}$ rapidly varies as $T^4$, and $c$ is the velocity of sound. 
There might be a numerical factor of order one in Eq.\,\eqref{Eq:2}, depending on the details of the phonon scattering (specular or diffuse). Because the wavelength of a thermal phonon in helium at 0.1 K is about 0.1 $\mu$m which is much smaller than the size of the sphere and likely also of the surface roughness, we expect both a geometric cross section and a diffuse scattering, but we are not aware of a calculation of the prefactor. We are assuming that it will be of order 1.   
Using the value of the radius $R=124\,\mu$m as determined optically, we obtain a perfect quantitative agreement with the data if we set this factor equal to one, see Fig.\,\ref{fig:2}. 
At temperatures below ca. 100 mK, the phonon drag falls below the background damping measured in vacuum. 
At temperatures above 1 K, the drag force is described by Stokes' solution extended to the two-fluid model \cite{Electronics,JaegerSchudererSchoepe1995,JaegerSchudererS1995a}.

\begin{figure}[h]
\centerline{\includegraphics[width=0.80\textwidth]{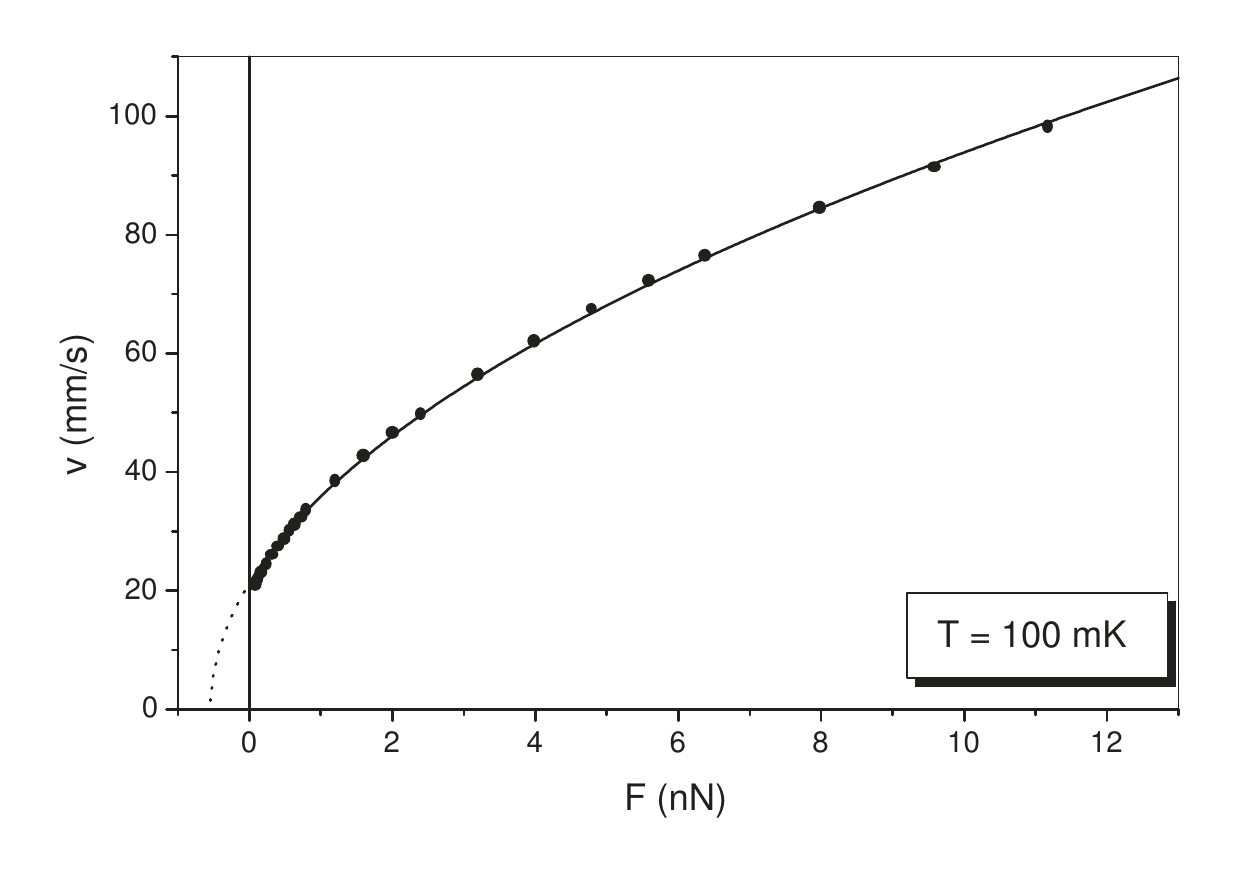}}
\caption{(From \cite{NiemetzSchoepe2004}) Velocity amplitude for turbulent flow at 100 mK at large driving forces. The solid line is the same parabolic fit as in Fig.\,\ref{fig:1}, see Eq.\,\eqref{Eq:4}.}  
\label{fig:3}
\end{figure}

We use this linear regime to determine the charge $q$ of the sphere. 
When a stable amplitude is reached at resonance, the drag force $\lambda v$ and driving force $F$ cancel. 
From the slope $(q/d)^2 /\lambda $ of the induced current $I$ as a function of $U_{ac}$, and from the time constant of the freely decaying amplitude $t_0\,=\,2m/\lambda$ measured separately at the same temperature, we obtain the charge (the mass $m$ and the spacing $d$ are known, see above). 
We apply this procedure at $300 \,\mbox{mK}$ where $t_0$ has a convenient value of $31\,\mbox{s}$. 
During the course of the experiment, we find the charge to remain constant, and occasionally a small loss of few percent can be observed after several weeks. 
\begin{bf}
\subsection{Stable turbulent flow}
\end{bf}
The nonlinear dependence of $v(F)$ in Fig.\,\ref{fig:1} and Fig.\,\ref{fig:3} can be properly described by a quadratic drag force. 
In contrast to a classical liquid, the apex of the parabolic shape of $v(F)$ is shifted to the left of the origin by 0.4 nN, and the resulting finite intercept at $F$ = 0 indicates a velocity range of frictionless flow which is the paradigm of superfluidity. 
Moreover, we observe a sharp onset of the turbulent regime, whereas in a classical liquid there are about three orders of magnitude in flow velocity between Stokes' regime of laminar flow and fully developed turbulence, where the classical drag on a sphere is given by $\gamma v^2$ with $\gamma = c_D \rho \pi R^2 / 2 $ ($\rho $ is the density of the liquid and the drag coefficient $c_D$ of a sphere is approximately 0.4).   

\begin{figure}[b]
\centerline{\includegraphics[width=0.80\textwidth]{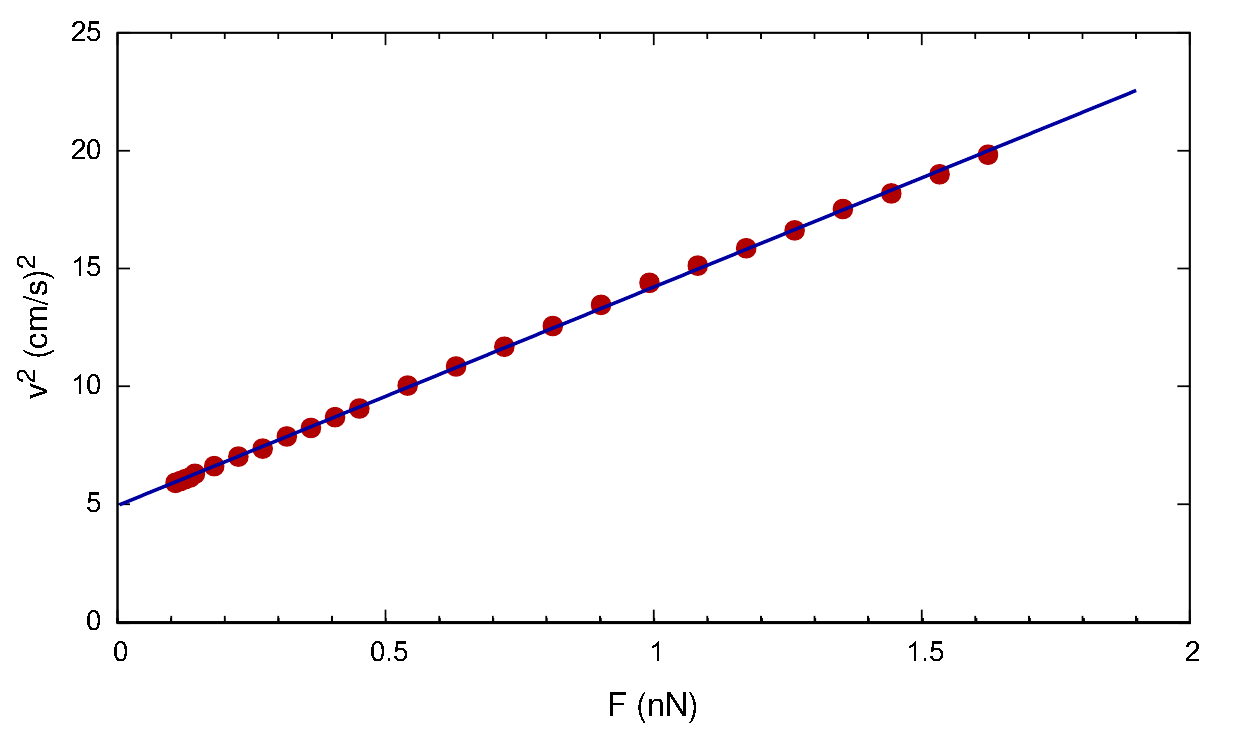}}
\caption{Turbulent flow at 33 mK, oscillation frequency 160 Hz. 
The straight line is a fit of a quadratic drag force, see Eq.\,\eqref{Eq:4}. From the slope we find a drag coefficient $c_D$ = 0.36. The intercept at $F = 0$ occurs at $v_c^2 \approx$ 5.0 (cm/s)$^2$. 
The background damping of the empty cell has been subtracted, phonon drag is negligible at 33 mK. (Color figure online)} 
\label{fig:4}
\end{figure}

For a quantitative analysis of the data, we have to take into account the following considerations. 
For oscillatory flow subject to a quadratic drag force, the concept of energy balance requires that per half-period the power injected by the drive and dissipation due to the drag must cancel for a stable oscillation amplitude. 
In the case of a linear drag, this leads to the simple result that the equilibrium amplitude is reached when driving force and drag force cancel, see above. 
But in the case of a quadratic drag of the form 

\begin{equation}
F_D = \gamma (v^2 - v_0^2) \geq 0\, , \label{Eq:3}
\end{equation}
that is obviously appropriate here, a short calculation gives for the driving force \cite{NiemetzKerscherSchoepe2001} 

\begin{equation}
F  = \frac{8\gamma}{3 \pi} ( v^2 - \frac{3}{2} v_0^2) \geq 0 \, .\label{Eq:4}
\end{equation}
There are 2 corrections, firstly a factor 8/3$\pi \approx$ 0.85, and secondly, an intercept for $F = 0$ at $v^2$ = 3/2 $v_0^2$. 
Thus, the intercept velocity is larger than $v_0$ by a factor $\sqrt {3/2} \approx  1.22$ . 
Before fitting Eq.\,\eqref{Eq:4} to the data, the linear drag (phonons and background), see Eq.\,\eqref{Eq:1}, must be subtracted. If we parametrize $\gamma$ like in the classical case, i.e., $\gamma = c_D\rho\pi R^2/2$, the only fitting parameter is the drag coefficient $c_D$ in $\gamma $. 
We find $c_D \approx$ 0.40 to within 10 \%. 
Hence, surprisingly, the expressions of $\gamma $ for oscillatory superflow and for classical uniform flow are identical. 
We {\it define} the critical velocity $v_c$ as the largest velocity at $F$ = 0, i.e., $v_c \equiv \sqrt{3/2}\,\, v_0$, which can be obtained either from the intercept in Fig.\,\ref{fig:4}, or directly from the kink in the $v(F)$ curve, see Fig.\,\ref{fig:5}.\\

\section{THE CRITICAL VELOCITY}\label{s.vc}

\begin{figure}[tb]
\centerline{\includegraphics[width=0.80\textwidth]{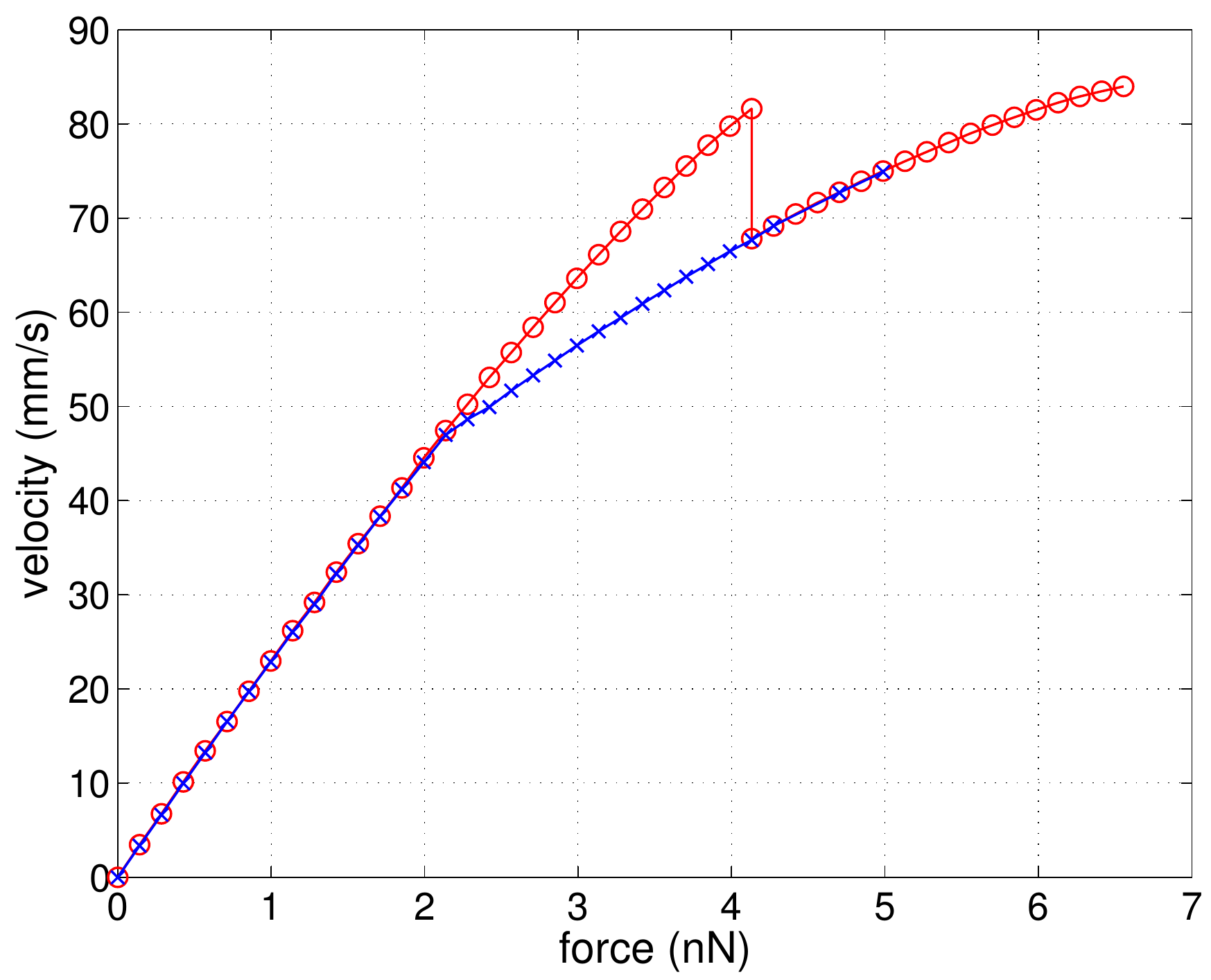}}
\caption{(From \cite{SchoepeHaenninenNiemetz2015}) Velocity amplitude as a function of the driving force at 1.9 K. 
Red circles: data taken with increasing drive; blue x: decreasing drive. 
Because of the strong hysteresis $v_c$ = 46 mm/s can only be determined with the down sweep. 
The oscillation frequency was 236\,Hz. (Color figure online)} 
\label{fig:5}
\end{figure}

When the velocity amplitude is increased beyond a certain value, the linear regime abruptly changes to a nonlinear regime at lower amplitude. 
This signals the transition to a large turbulent drag force that is followed by a hysteresis with the down sweep, see Fig.\,\ref{fig:5}.

We interpret the hysteretic behavior as follows. 
At the beginning of the first up-sweeps there are few remanent vortices left from the first cool-down. 
Potential flow will break down when the oscillating sphere collides with such a vortex and then sheds more vortices. 
In fact, detailed experiments with vibrating wires, e.g., by the Osaka group, demonstrate that remanent vorticity can be reduced, or even avoided, if the measuring cell is filled very slowly at low temperatures \cite{Yano}. 
In that case, no transition to turbulence was observable up to very high velocity amplitudes of 1.5 m/s. Obviously, the wire does not shed vortices by itself. 
We expect the same scenario to occur in our case. 
From Fig.\,\ref{fig:5} the  velocity at initial breakdown is $v_c^*$ = 82 mm/s and the critical velocity is $v_c$ = 46 mm/s. 
We may infer the intervortex spacing $l_0 = L_0^{-1/2}$ where $L_0$ is the remanent vortex density (defined as the length of all vortices per unit volume, having a dimension 1/m$^2$) by postulating that the critical oscillation amplitude is determined by the spacing of the remanent vortices, i.e.,   
\begin{equation}
\frac{l_c}{l_0} = \frac{v_c}{v_c^*} \, . \label{Eq:5}
\end{equation}
We find a ratio of 0.56 for the spacings and hence a ratio $L_0 / L_c$ = 0.31. 
If for some reason $L_0$  were too large, a too small $v_c$ would be measured. 
This and the hysteresis must be taken into account when determining $v_c$. 

\begin{figure}[t]
\centerline{\includegraphics[width=0.75\textwidth]{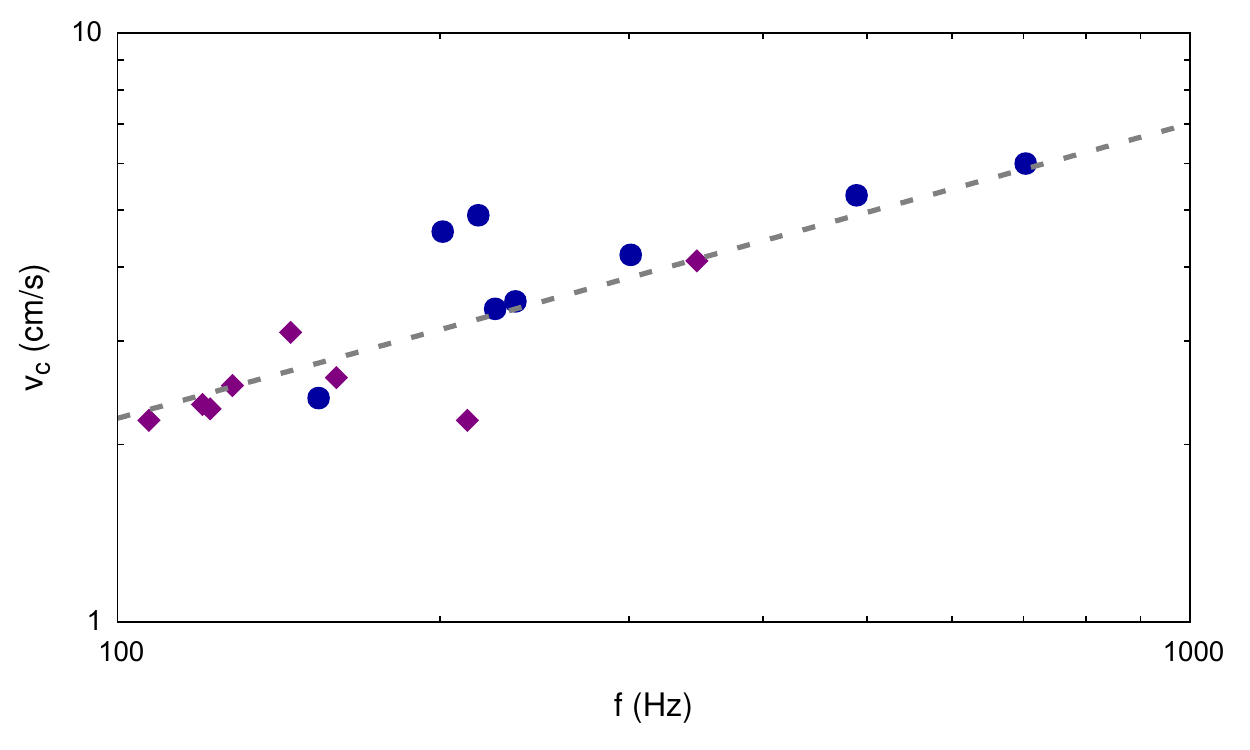}}
\caption{(From \cite{HaenninenSchoepe2008}) Critical velocity for the onset of turbulence as a function of the frequency of 2 oscillating spheres. 
Violet diamonds: radius 124\,$\mu$m; blue dots: 100\,$\mu $m; black line: calculated from Eq.\,\eqref{Eq:6}. 
There is a large scatter of the data near 200 Hz that might be caused by some unusual remanent vorticity. (Color figure online)} 
\label{fig:6}
\end{figure}

We now turn to dependence of $v_c$ on the oscillation frequency $\omega = 2 \pi f$. 
Empirically we find a scaling of $v_c$ as $\omega ^{1/2}$, see Fig.\,\ref{fig:6}, in agreement with results obtained with a vibrating wire at Osaka \cite{Yano2} and with a tuning fork at Lancaster \cite{Pick}. 
On dimensional grounds this leads to $v_c \sim \sqrt{\kappa \, \omega} $. 
Because there is still no rigorous theory available the numerical prefactor must be determined from the experiment. 
In our case, we find from Fig.\,\ref{fig:6}
\begin{equation}
v_c \approx 2.8 \sqrt{\kappa \, \omega} \, .\label{Eq:6} 
\end{equation}\\
The other authors, working with vibrating wires or tuning forks, find slightly different numerical prefactors, both are a little smaller but also of order 1. 
The shape of the oscillating object might have an effect. 
For example, sharp corners strongly increase the velocity locally, making the prefactor smaller.

Although we have no theory for oscillating superfluid flow, there are several arguments that make the $\omega ^{1/2}$ dependence of $v_c$ plausible \cite{HaenninenSchoepe2008a}. 
Firstly, starting from the superfluid Reynolds number $Re_s \equiv v\,D/\kappa \sim $ 1 for the onset of turbulence, and choosing the oscillation amplitude as the characteristic length scale $D = v/\omega $ we obtain $v_c \sim \sqrt{\kappa \,\omega} $. 
This is a qualitative, but very general argument. 
Note that the size $R$ of the oscillating object is irrelevant here in contrast to the Feynman critical velocity for uniform motion $\kappa /R$, where the size is the relevant length scale $D = R$.

\begin{figure}[t]
\centerline{\includegraphics[width=0.80\textwidth]{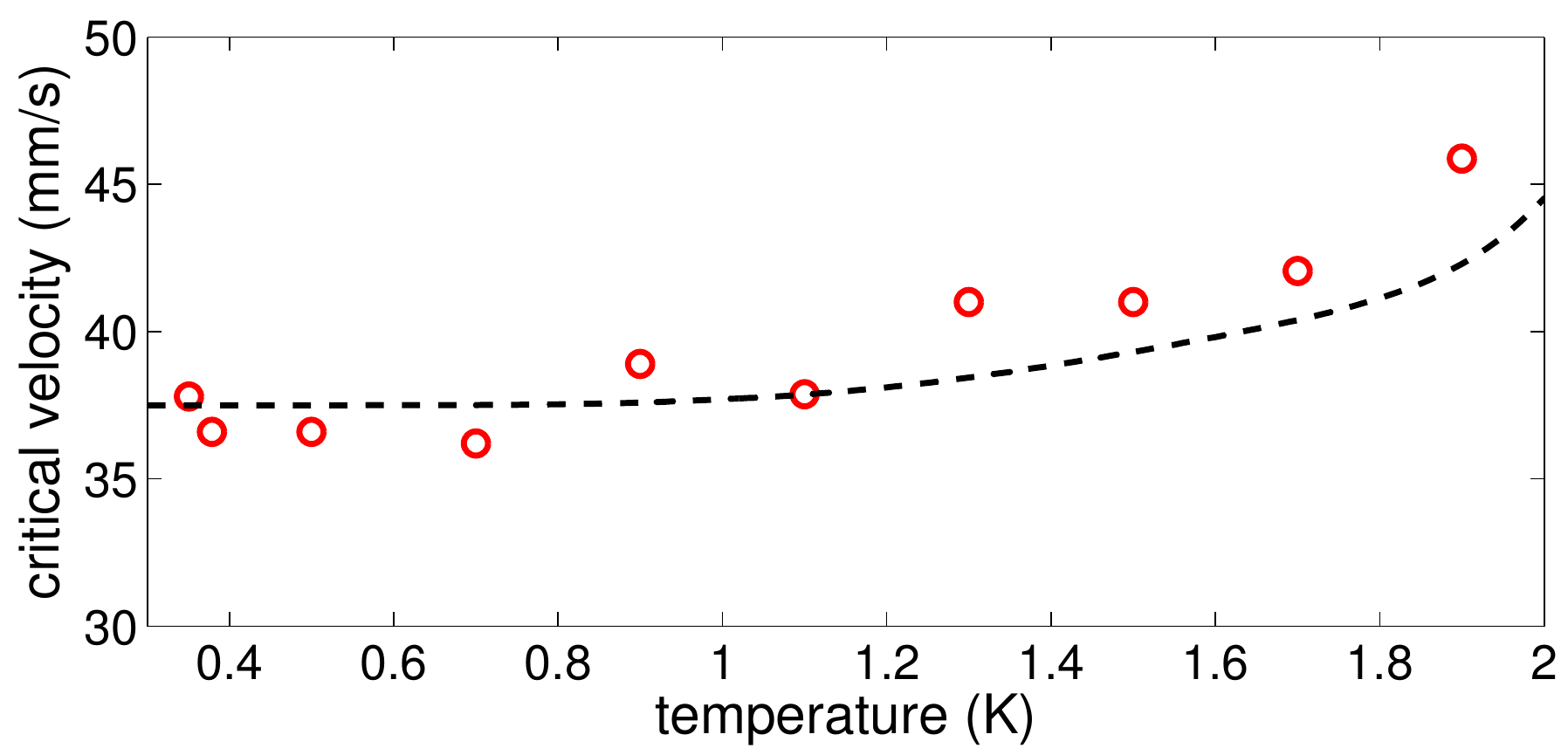}}
\caption{(From \cite{HaenninenSchoepe2008a}) Temperature dependence of the critical velocity (oscillation frequency 236 Hz). 
The variation due to $\beta (T)$ is indicated by the dashed line. (Color figure online)} 
\label{fig:7}
\end{figure}

Secondly, making use of results obtained by Kopnin \cite{Kopnin} based on the well known Vinen equation for uniform motion, a relaxation time $t_1 $ of the vortex tangle after a change of the superfluid flow velocity $v_s$ has been calculated: 

\begin{equation}
t_1 = 2 \kappa /\beta v_s^2 \, . \label{Eq:7}
\end{equation}
Here the numerical parameter $\beta $ = A(1 - $\alpha ^{\prime} $) - B$\alpha $ is determined by the coefficients of mutual friction $\alpha ^{\prime} $ and $\alpha $ and the constants A,\,B $\sim $ 1. 
Applying this to oscillatory flow (which remains to be proven valid), we take the time the velocity changes from zero to the maximum, i.e., one quarter of a period, in which the vortex tangle can follow and thus arrive at the condition
\begin{equation}
\omega t_1 < 1/4 \, , \label{Eq:8}
\end{equation} 
and therefore we have

\begin{equation}
v_s \ge  v_c \approx \sqrt{8\,\kappa \,\omega /\beta } = 2.83 \sqrt{\kappa \,\omega /\beta }\, . \label{Eq:9}
\end{equation}\\
To compare Eq.\,\eqref{Eq:9} with the experimental data we evaluate $\beta $ for $^4$He by referring to tabulated values of $\alpha$ and $\alpha^{\prime}$ \cite{alpha}. 
Setting the constants A,B = 1, we find $\beta $ = 0.95 (at 1.3\,K), 0.89 (at 1.6\,K), and 0.79 (at 1.9\,K). 
Below 1\,K mutual friction is very small, hence we set $\beta = 1$. 
This implies a slow increase of $v_c$ by about 10\% in qualitative agreement with experimental results as displayed in Fig.\,\ref{fig:7}. 
Closer to the $\lambda $-point, the critical velocity could not be identified accurately any more because there was no sharp onset of turbulence detectable in the $v(F)$ curves.\footnote{An extrapolation of the linear and the turbulent part of $v(F)$, e.g., in a double-logarithmic plot, indicates a velocity above which the turbulent drag dominates the linear drag, but not the {\it onset} of turbulent drag, hence not $v_c$ that actually could be much smaller \cite{Jack}.}
 
\begin{figure}[tb]
\centerline{\includegraphics[width=0.70\textwidth]{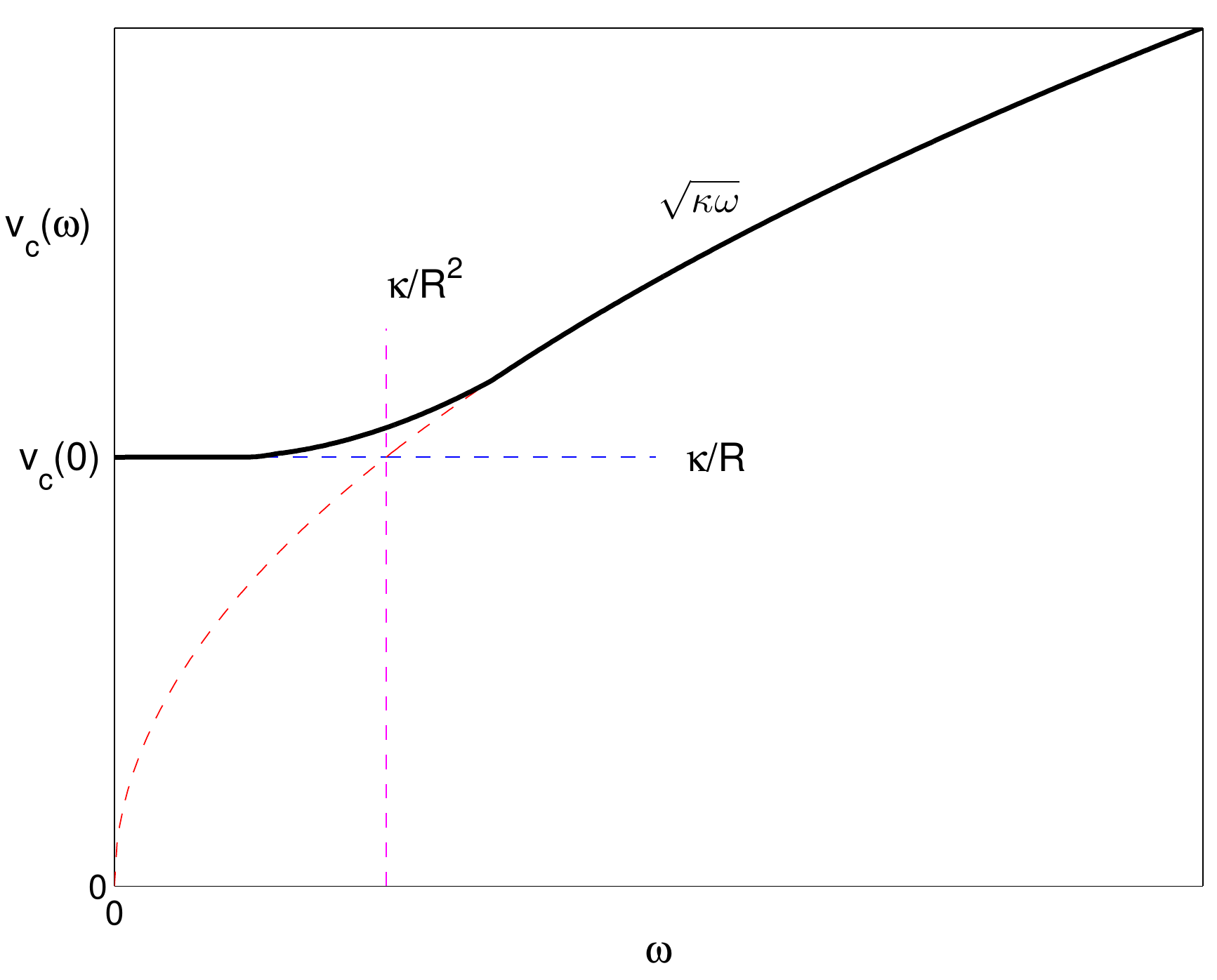}}
\caption{(From \cite{HaenninenSchoepe2010}) Sketch of the crossover of the critical velocity $v_c (\omega) $ from oscillatory flow to steady flow $v_c(0)$ at $\omega \sim \kappa /R^2$. (Color figure online)} 
\label{fig:8}
\end{figure}

At present, the most accurate theoretical support of the $\omega \,^{1/2}$ dependence of $v_c$ comes from a dynamical scaling of the critical velocity \cite{HaenninenSchoepe2011}. 
Instead of presenting here all details of this approach, we only outline the following concept. 
The equation of motion of a vortex filament was shown by Schwarz \cite{Schwarz} to be invariant when scaling the length $l = \delta \, l^*$, the velocity $v = v^*/\,\delta  $, and the time $t = \delta ^2\,t^*$. 
This result was extended to oscillatory motion by Kotsubo and Swift \cite{Swift} by recognizing that the frequency $\omega $ is a scaling variable. 
	The final result, that is exact to within a small logarithmic correction of about 4\% in our frequency range $\omega ^*/\omega  \approx $ 7, is

\begin{equation}
v_c/v_c^* \approx (\omega /\omega ^*)^{1/2} \, .\label{Eq:10}
\end{equation}
\\
It is quite clear that the $\omega \,^{1/2}$ dependence of $v_c$ cannot remain valid when the frequency approaches zero for uniform motion, because a finite critical velocity is known to exist in that case too. 
We are suggesting that the relevant length scale in this case is the size $R$ of the sphere. 
Thus, as shown above, we recover the Feynman critical velocity $v_c(0) \sim \kappa /R$ = 0.5 mm/s. 
The transition to steady flow will occur when $\sqrt{\kappa \omega } \sim \kappa /R$, that is at $\omega \sim \kappa /R^2$ = 6.5 s$^{-1}$, which is much smaller than our oscillation frequencies ($>$ 700 s$^{-1}$), see Fig.\,\ref{fig:8}. 
For much smaller objects this will matter.

Finally, we should mention that with our experimental method, we have no quantitative control of the resonance frequency of the sphere. 
It depends on the state of levitation, in particular on how much flux is trapped in the superconducting electrodes. 
A more rapid cool down in our $^3$He cryostat in general produced higher frequencies than in our dilution cryostat.

\section{\,INTERMITTENT SWITCHING BETWEEN POTENTIAL FLOW AND TURBULENCE}
Below $0.5\,\mbox{K}$ the hysteretic regime discussed in the previous section is replaced by an instability where neither potential flow nor turbulence are stable, but instead the flow switches intermittently between both patterns  (shaded area in Fig.\,\ref{fig:1}). The unstable regime extends from the critical velocity $v_c$ up to $v_c + \Delta v $ where $\Delta v /v_c \le$ 3\%. It is plausible that stable turbulence cannot exist exactly at $v_c$, because in the limit $T\rightarrow0$ no power is available at vanishing drive to create vorticity. 

\begin{figure}[h]
\centerline{\includegraphics[width=0.8\textwidth]{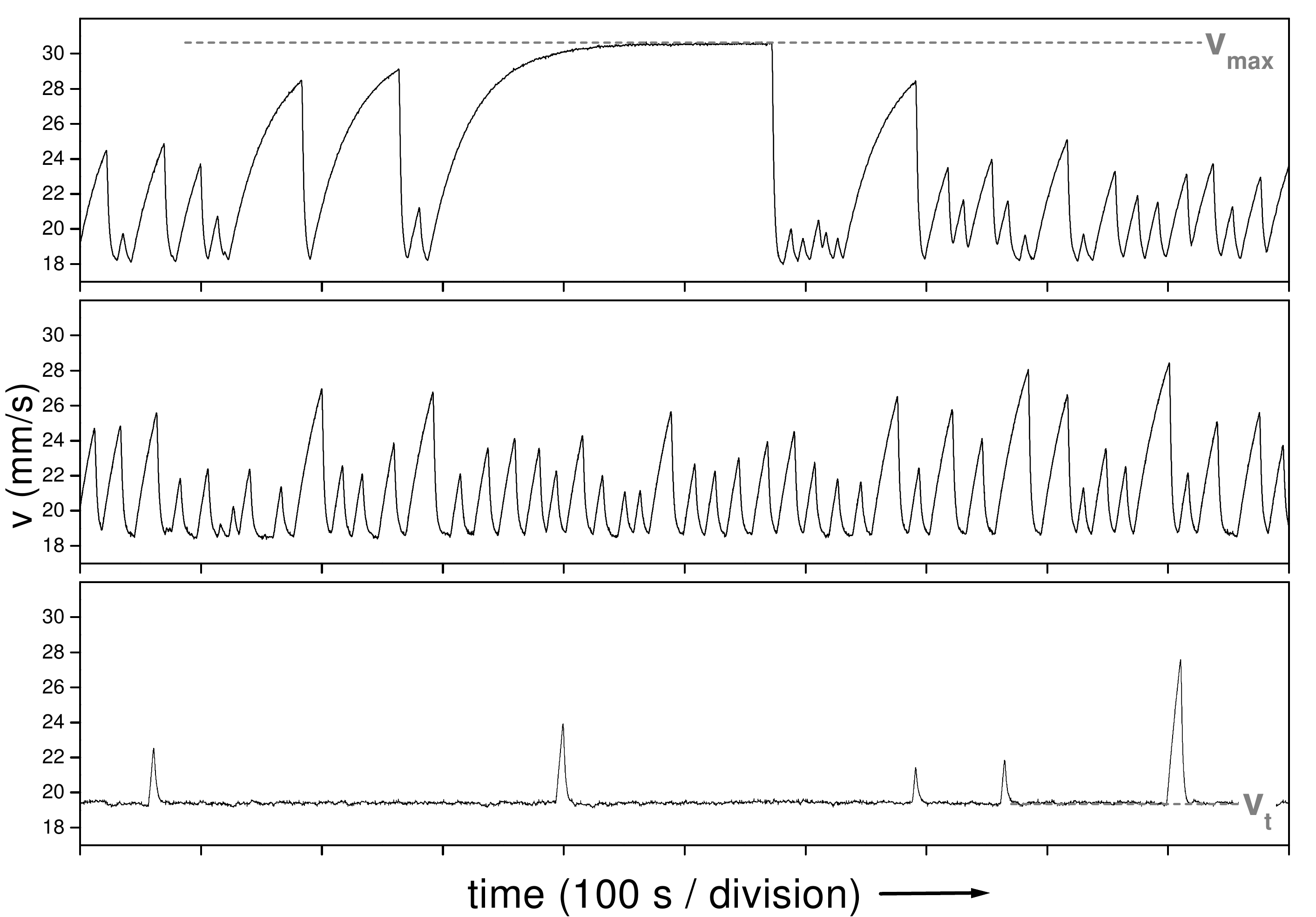}}
\caption{(From \cite{NiemetzKerscherSchoepe2002}) Three time series of the velocity amplitude at 300 mK and at three different driving forces (in pN: 47, 55, and 75), are shown from top to bottom. The low level $v_t$ corresponds to turbulent flow while the increase occurs during a laminar phase. With increasing drive the lifetimes of the laminar phases become shorter, whereas the lifetimes of turbulent phases grow rapidly. The time interval shown here extends over 1000 s $\approx $ 17 min.}
\label{fig:9}
\end{figure}

We have recorded time series (lasting up to 36 hours)  
of this switching phenomenon at various constant driving forces and temperatures, see Fig.\,\ref{fig:9}, for 
three different driving forces at $300\,\mbox{mK}$. The amplitude 
switches between a low level $v_t$ corresponding to turbulent drag and an 
exponential recovery of the laminar level corresponding to phonon 
scattering. Because the velocity amplitudes are above $v_c$, these phases of potential flow break down at some velocity $v_c^* > v_c$. If the laminar phase lasts long enough, the stationary value for laminar flow $v_{\mathsf{max}}$ is reached (see upper time series). When the laminar flow 
breaks down, a rapid drop to the turbulent level occurs due to the onset of the large and nonlinear drag. It is obvious from Fig.\,\ref{fig:9}, that with increasing driving force the lifetimes of the 
laminar phases become shorter and those of the turbulent phases grow very rapidly. In the following we analyze these time series statistically. 
\begin{bf}
\subsection{Stability of the laminar phases}
\end{bf}
The statistical properties of the laminar phases \cite{NiemetzKerscherSchoepe2002} can be analyzed either from the 
distribution of the signal height $\Delta v = v_c^* - v_t$ (velocity at breakdown minus 
turbulent velocity $v_t$), or by considering the distribution of the 
lifetimes. Both variables are related by the exponential recovery of the laminar amplitude (starting at $t=0$) 
\begin{equation}
\Delta v(t) = \Delta v_{\mathsf{max}} (1-\exp(-t /t_0))\, , \label{Eq:11}
\end{equation}
where $\Delta v_{\mathsf{max}}$ 
is the difference between the stationary laminar velocity amplitude 
$v_{\mathsf{max}}=F/\lambda$ and $v_t$, and $t_0 = 2m/\lambda$ is the time constant.

\begin{figure}[h]
\centerline{\includegraphics[width=0.80\textwidth]{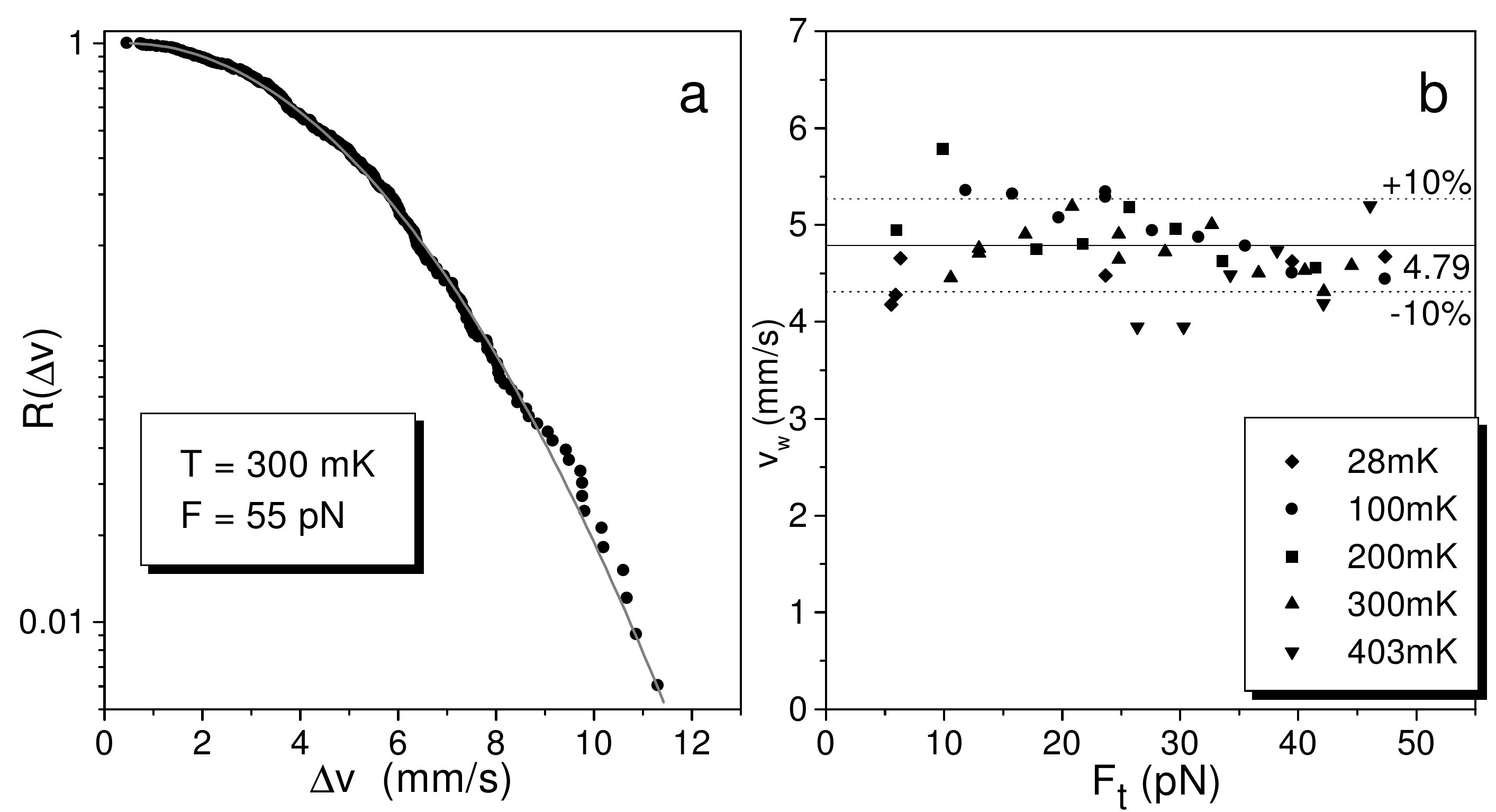}}
\caption{(From \cite{NiemetzKerscherSchoepe2002}) Statistical analysis of the laminar phases. a) The Rayleigh distribution $R(\Delta v)\,=\,\exp(-(\Delta v/v_w)^2)$ of the velocity increase $\Delta v$ above the turbulent level. b) The fitting parameter {$v_w$} is shown to be independent of temperature and driving force $F_t = F - \lambda v_t$, i.e., linear drag force subtracted.}
\label{fig:10}
\end{figure}

In Fig.\,\ref{fig:10}a the distribution of the signal height $\Delta v$ is 
shown to be of the Gaussian form 
\begin{equation}
R(\Delta v) = \exp(-(\Delta v/v_w)^2)\, , \label{Eq:12}
\end{equation} 
where $v_w$ is a fitting parameter. The distribution function describes the probability $R(\Delta v)$ that a given $\Delta v$ is exceeded (``reliability function"). In our case, it is the so-called Rayleigh distribution, and the fitting parameter $v_w \approx 4.8\,\mbox{mm/s}$ (resonance frequency 120 Hz) is found to be completely independent of both temperature and driving force, see Fig.\,\ref{fig:10}b. It is a property of the Rayleigh distribution that $v_w$ is the rms value of the distribution. 

From the Rayleigh distribution we obtain the failure rate $\Lambda (\Delta v)$ which is the conditional probability that the potential flow decays in a small interval just above $\Delta v$, provided it has not decayed until $\Delta v$:

\begin{equation}
\Lambda(\Delta v)=-\frac{d\ln R}{d\Delta v}= \frac{2\Delta v}{ v^2_w} \, . \label{Eq:13}
\end{equation}

We note that the failure rate is proportional to $\Delta v$, and hence, to the increase of the oscillation amplitude $\Delta a = \Delta v / \omega$. This fits into our picture that a collision of the sphere with remanent vortices will cause a breakdown of the potential flow, see Fig.\,\ref{fig:5}, where the breakdown at $v_c^* > v_c$ is related to the intervortex spacing $l_0$ of the remanent vortex density.

Another property of the failure rate can be found when changing the variable from $\Delta v$ to the lifetime $t$ of the potential flow:

\begin{equation}
\Lambda(t)=\Lambda(\Delta v(t)) \cdot \frac{d\Delta v}{dt} = \frac{2 \Delta v(t)}{v_w^2}  \cdot \frac{d\Delta v}{dt}=\frac{1}{v^2_w}\frac{d(\Delta v)^2}{dt}\, ,\label{Eq:14} 
\end{equation} 
where $\Delta v(t)$ is given by Eq.\,\eqref{Eq:11}. 
Initially we have a linear increase of $\Delta v(t)$ and therefore $\Lambda(0) = 0$. At a time $t$ = $t_0\ln{2}$ the failure rate has a maximum. And, if the maximum level $\Delta v_{\mathsf{max}}$ is approached for large $t$, we find $\Lambda  \rightarrow 0$. Thus, the flow becomes stable, although the velocity $v_{\mathsf{max}}$ is clearly above $v_c$.

\begin{figure}[t]
\centerline{\includegraphics[width=0.75\textwidth]{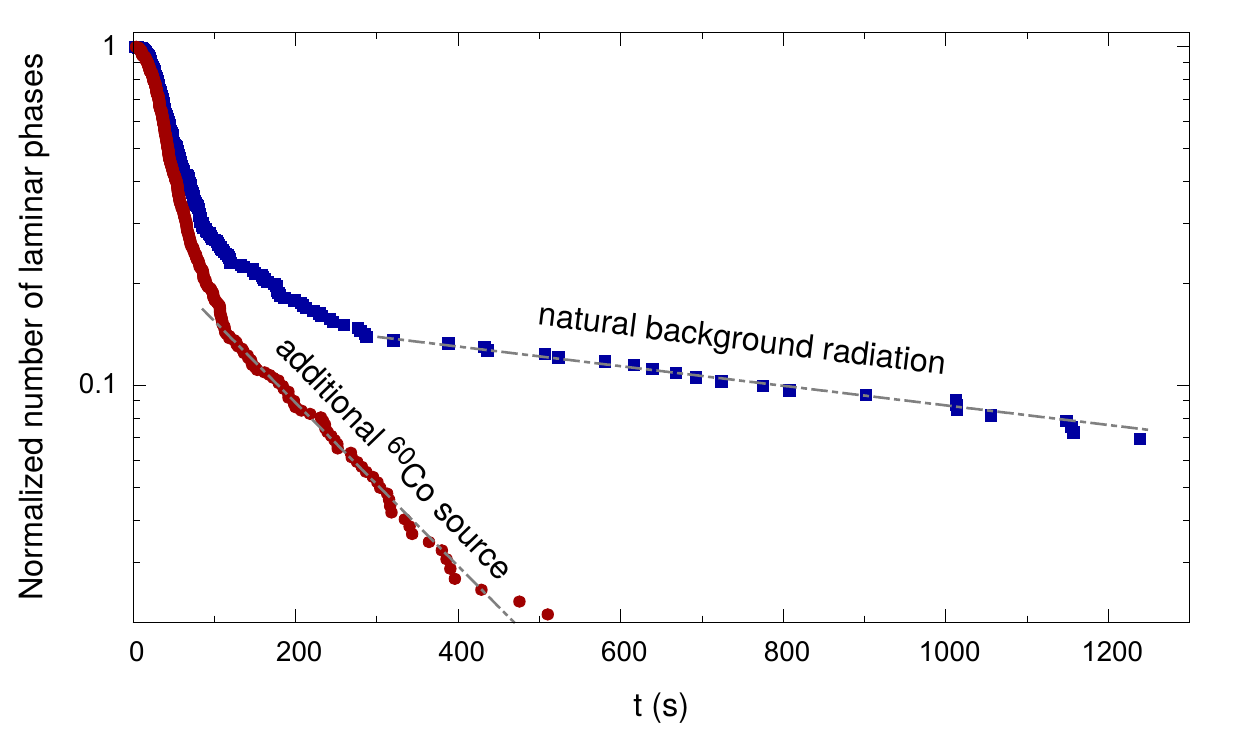}}
\caption{Distribution of the lifetimes of the long-lived laminar phases slightly above the 
critical velocity $v_c$. For times $t \gg t_0 = 31\,\mbox{s}$ (at 300 \,mK) a 
mean lifetime of 25 minutes is measured. The additional radioactive source reduces the mean lifetime to 3.0 
minutes. (Color figure online)}
\label{fig:11}
\end{figure}

These metastable states of potential flow above $v_c$ will also break down after a mean lifetime of 25 min, because natural radioactivity produces vorticity in the fluid. This is proven by placing a weak radioactive source $^{60}\mathrm{Co}$ (activity of $74\,\mbox{kBq}$ = 2 $\mu$Ci) near the cryostat, resulting in a dramatic effect: the mean lifetime of the laminar phase is 
reduced to 3.0 minutes, see Fig.\,\ref{fig:11}.

We have measured the dose rate of the source at the 
position of the measuring cell inside the cryostat (taking into account a 
measured 20\% loss in the dewar walls) to be 440\,nGy/h ($\pm$\,5\%). 
Comparing this value with the measured dose rate due to natural background 
radiation in our laboratory of 57.6 ($\pm$ 1.5)\,nGy/h , 
we obtain an increase of the dose rate due to the source by a factor of $(440+57.6)/57.6 = 8.64$. Within the error bars this compares well with 
the measured reduction of the mean lifetime of the laminar phases by a 
factor of $25/3.0=8.3$.

The effect of the radioactivity is attributed to the creation of He ions in the superfluid, 
which produce local vorticity either during the creation and recombination 
processes, or when the ions are accelerated by the electric field which 
exists in our measuring cell. This vorticity then triggers the 
breakdown of the potential flow when the velocity amplitude of the sphere reaches 
the corresponding critical velocity $v_c^*$. We therefore conclude that the 
lifetime of the metastable laminar phases which we observe above $v_c$ is 
limited only by natural background radioactivity. 
\\

Below 100 mK the phonon drag is negligibly small, only the residual damping of the empty cell $\lambda_{\mathsf{res}}$ = 4.4 $10^{-11}$ kg/s determines the oscillation amplitude at a given drive. Equivalent is a time constant $t_0$ = 2m/$\lambda_{\mathsf{res}}$ = 1200 s. Therefore, the exponential function in Eq.\,\eqref{Eq:11} can be expanded:

\begin{equation}
\Delta v(t) = \left(\frac{F}{\lambda_{\mathsf{res}}} - v_t\right) \frac{t}{t_0} = \left(\frac{F}{2m} - \frac{v_t\lambda_{\mathsf{res}}}{2m}\right)t\,. \label{Eq:15}
\end{equation}

\begin{figure}[!h]
\centerline{\includegraphics[width=0.8\textwidth,clip=true]{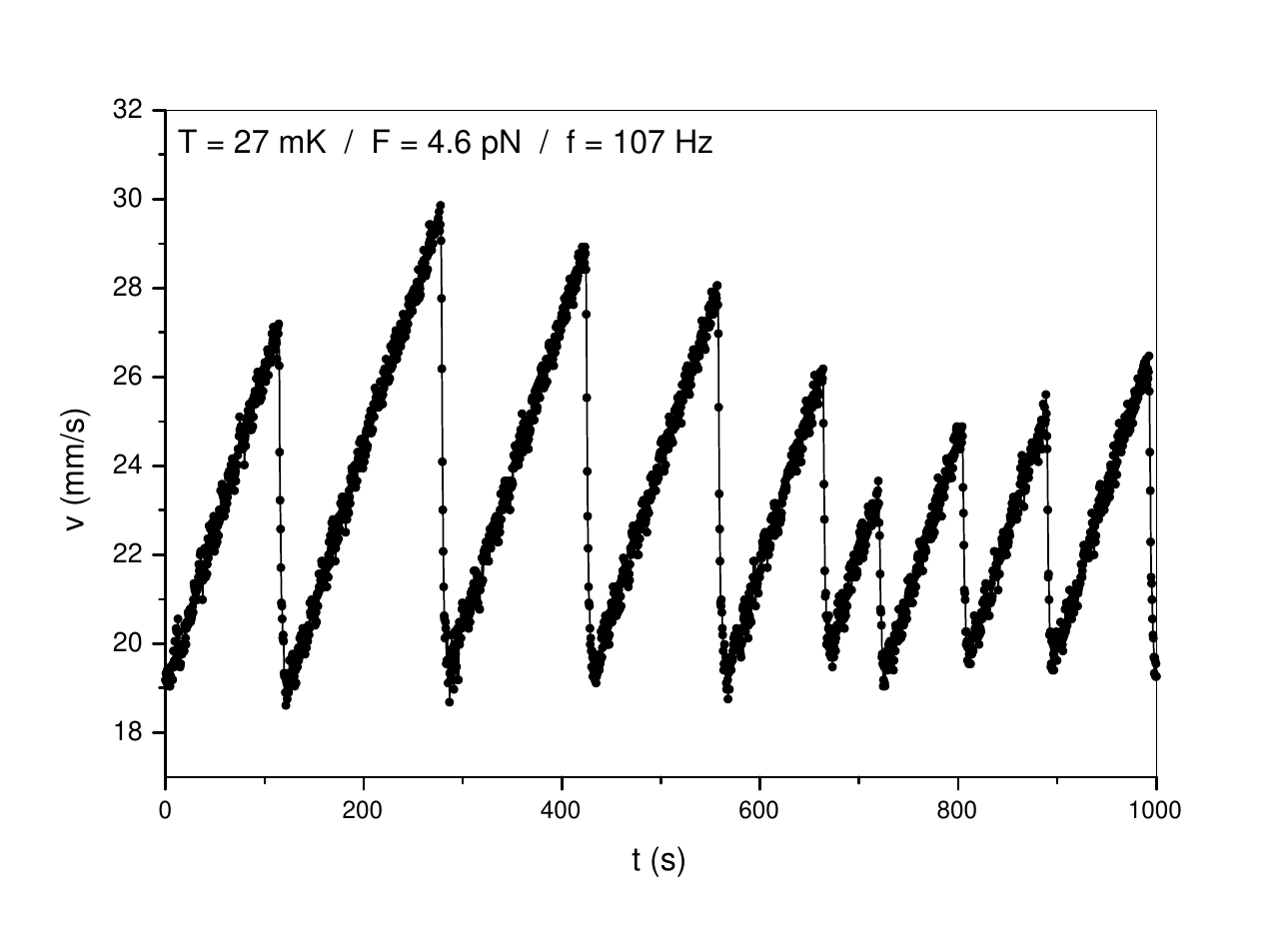}}
\caption{(From \cite{SchoepeHaenninenNiemetz2015}) Time series of the breakdown of the phases of potential flow recorded at 27 mK with a driving force of 4.6 pN. The linear increase of $v(t)$ during potential flow is described by Eq.\,\eqref{Eq:15}, see text. The lifetimes of the turbulent phases cannot be resolved (not even on an expanded time scale). We take the minima near 19 mm/s as $v_t \approx v_c$.                    }
\label{fig:12} 
\end{figure}

In Fig.\,\ref{fig:12} the time series recorded at 27 mK with a driving force of 4.6 pN shows a stochastic saw tooth pattern of the laminar phases. The lifetimes of the turbulent phases cannot be resolved any more. %The smallest velocity of a stable turbulent flow was measured to be 19.6 mm/s with a drive of 55 pN. Therefore, we assume that the base line in Fig.\,\ref{fig:12} at 19 mm/s can safely be taken as $v_c$. 

\begin{figure}[h]
\centerline{\includegraphics[width=0.7\textwidth,clip=true]{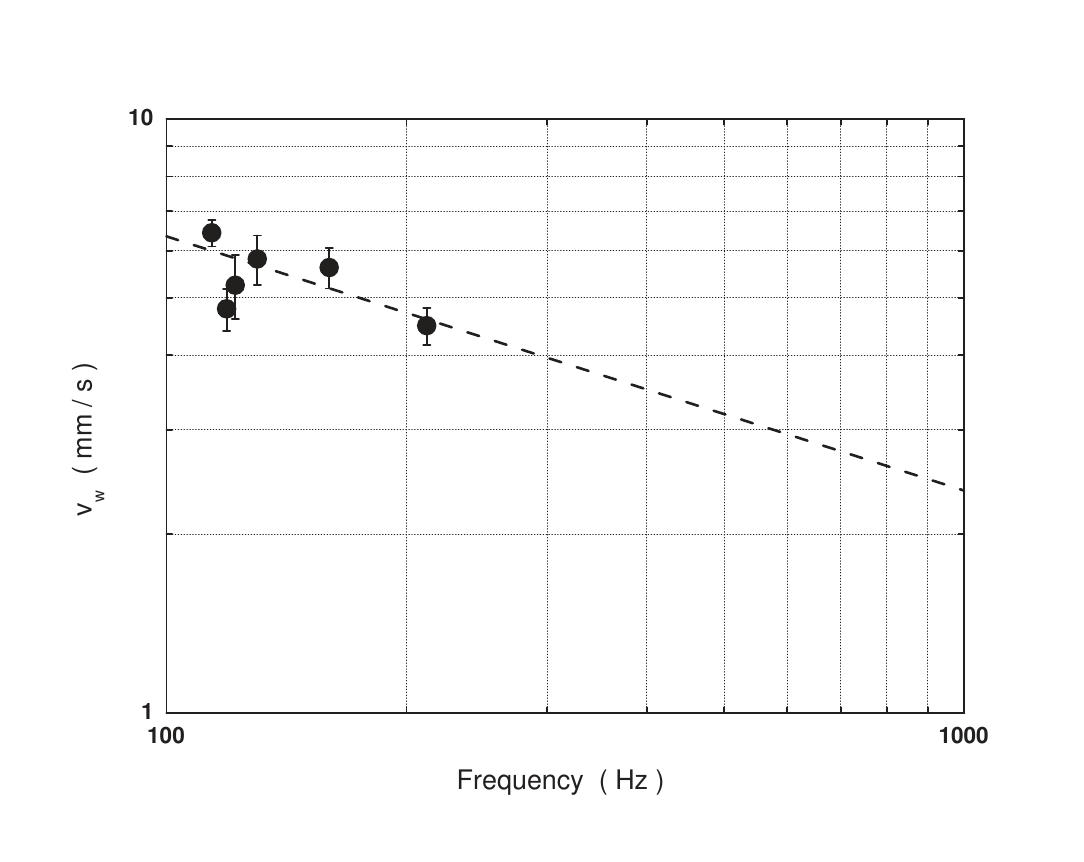}}
\caption{(From \cite{SchoepeHaenninenNiemetz2015}) The rms excess velocities $v_w$ for six different frequencies. Each data point is obtained from the average of several individual time series recorded at different temperatures and driving forces. The error bars represent the standard deviations. A power law fit that takes into account the error bars indicates a slope of $-0.43 \pm 0.14$.}
\label{fig:13}       
\end{figure}

The slope of the linear increase of the velocity amplitude during potential flow amounts to 6.9 $10^{-5}$ m/s$^2$. Inserting in Eq.\,\eqref{Eq:15} the relevant numbers for $F$ = 4.6 pN, $v_t \approx $ 19 mm/s, $m$ = 27 $\mu$g, and $\lambda_{\mathsf{res}}$ = 4.4 $10^{-11}$ kg/s, we find a slope of  7.0 $10^{-5}$ m/s$^2$, in agreement with the experimental value. From Eq.\,\eqref{Eq:15} follows that the lifetimes of the phases of potential flow now also have a Rayleigh distribution. From the measured rms value $v_w$ = 6.5 mm/s (resonance frequency 107 Hz) of the $\Delta v$ distribution, we obtain the rms lifetime $t_w = 2 m v_w/ (F - v_t\lambda_{\mathsf{res}})$ = 93 s. 
The second term in the brackets of the rhs of Eq.\,\eqref{Eq:15} which is due to the residual damping, is here 18\% of the first one, and decreases further at larger drives. Finally, only the first term will dominate, i.e., we may neglect damping and approach the case of an undamped oscillator:  

\begin{equation}
\Delta v(t) = \frac{Ft}{2m}\, . \label{Eq:16}
\end{equation}
The rms lifetime $t_w$ of the potential flow is then proportional to $1/F$:
\begin{equation}
t_w = 2 m v_w/ F\, , \label{Eq:17}
\end{equation}
and the failure rate, Eq.\,\eqref{Eq:14}, would simply be given by
\begin{equation}
\Lambda(t) = \left(\frac{F}{2 m v_w}\right )^2 2\,t = \frac{2\,t}{t_w^2} \, . \label{Eq:18}
\end{equation}
We observe a weak frequency dependence of $v_w$, see Fig.\,\ref{fig:13}.

A power law fit to the data indicates a slope of $- 0.43 \pm 0.14$. This may be interpreted as an $\omega ^{-1/2}$ law. However, our frequency range is rather limited ($\sim$ factor 2) and there is some scatter. Therefore, this dependence is not firmly established. However, the data point at 212 Hz is the mean of 14 individual $v_w$ data measured at different temperatures (28 mK and 300 mK) and 11 different driving forces, and therefore cannot simply be neglected. A theoretical interpretation is not possible as long as a theory of the transition to turbulence in oscillating superflows is not available. However, a strong support of this frequency dependence of $v_w$ comes from the failure rate Eq.\,\eqref{Eq:18}. In that case, it follows that $\Lambda (t) \propto \omega \, t$, i.e., the failure rate is proportional to the number of the completed cycles. This appears to be plausible.

The analysis of our data allows us to obtain informations concerning the properties of the unstable phases of potential flow above the critical velocity. The failure rate Eq.\,\eqref{Eq:13} is proportional to the increase of the oscillation amplitude beyond $v_c$, and is independent of temperature and driving force. The final breakdown of potential flow at $v_c^* > v_c$ is known to be affected by the remanent vorticity and occurs statistically during the switching process at mK temperatures. Moreover, the stability of the phases of potential flow when $v_{\mathsf{max}}$ is reached (see upper trace in Fig.\,\ref{fig:9}) leads us to assume that in this case the intervortex spacing $l_0 = L_0^{-1/2}$ of the remanent vorticity $L_0$ cannot be reached by the sphere: $l_0 > v_{\mathsf{max}}/\omega \approx$ 40 $\mu$m (less than half of the radius of the sphere).

Defining $\Delta l = l_0 - l_c $ we obtain from Eq.\,\eqref{Eq:5} the relation

\begin{equation}
\frac{\Delta v}{v_c} = \frac{\Delta l}{l_c}\, . \label{Eq:19}
\end{equation}
The rms value $v_w$ of the $\Delta v$ distribution is now related to the rms value $l_w$ of the $\Delta l$ distribution:

\begin{equation}
\frac{v_w}{v_c} = \frac{l_w}{l_c} \, .\label{Eq:20}
\end{equation}
Inserting $v_w$ = 6.5 mm/s and $v_c$ = 19 mm/s (data from Fig.\,\ref{fig:12}), we get $l_w/l_c$ = 0.34 and thus an average $l_0$ = 1.34 $l_c$ or an average $L_0/L_c$ = 0.56. This means that on the average 56 \% of the vortex density at $v_c$ have survived the breakdown of the turbulent phase. But as long as the value $L_c$ is unknown for oscillatory flow, we cannot determine absolute values of the vortex densities. Also, since our sphere radius $R$ may be larger than the intervortex distance $l_0$, it is rather peculiar that the transition to turbulence is not triggered by the vortices likely to be attached to the sphere. However, it is possible that such vortices are absent, and in the laminar state the surface of the sphere becomes free from vorticity, or is only covered by very tiny loops which require a large velocity in order to expand. The transition to turbulence can then occur at oscillation amplitudes that are smaller than $R$. This would happen when the sphere comes again in contact with the tangle that still remains in the vicinity of the sphere. The correct physical picture requires experiments that could visualize the vortices around the sphere or realistic computer simulations.

We understand now why intermittent switching is not observed above 0.5 K: because of the large phonon drag, the velocity increase $\Delta v_{\mathsf{max}}$ is rather small and so is the failure rate. Only by increasing the drive, the laminar phase will finally break down and the turbulent phase will be stable. Only a considerable reduction of the drive will cause a transition back to potential flow. Hence, instead of intermittent switching, a hysteresis is observed at higher temperatures. 

Hysteresis and switching of the flow have been observed also with vibrating wires \cite{Yano3} and tuning forks \cite{Pick2}, and the significance of remanent vorticity for the critical velocity $v_c^*$ at breakdown was demonstrated. In our work we are relating $v_c^*$ directly to the intervortex spacing of the remanent vortex density.\\

\begin{bf}
\subsection{Lifetimes of turbulent phases}
\end{bf}

\begin{figure}
\begin{center}
\includegraphics[width=.80\textwidth]{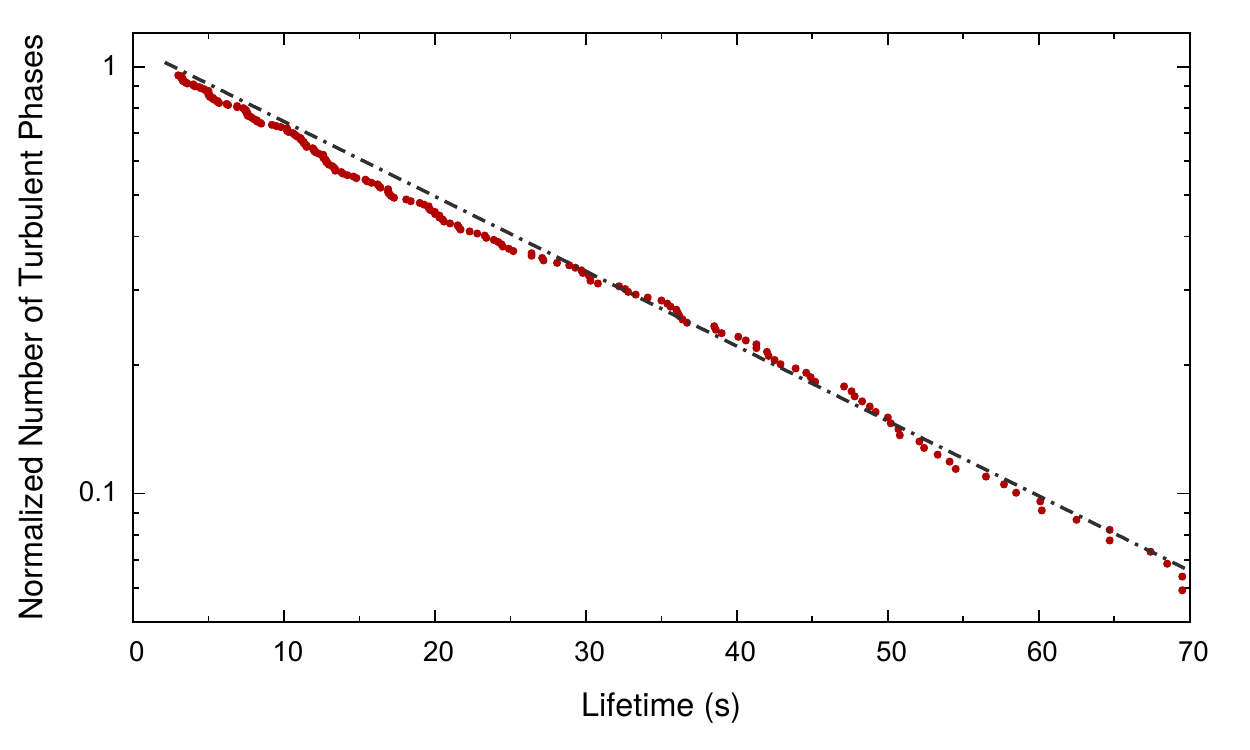}
\end{center}
\caption{ Exponential distribution of the lifetimes of the turbulent phases in a time series of about 2 hours,  recorded at 300 mK. The mean lifetime is $\tau $ = 24.8 s. Very short lifetimes (below ca. 1 s) could not be resolved. (Color figure online)}
\label{fig:14}       % Give a unique label
\end{figure}

\begin{figure}[b]
\begin{center}
\includegraphics[width=.90\textwidth]{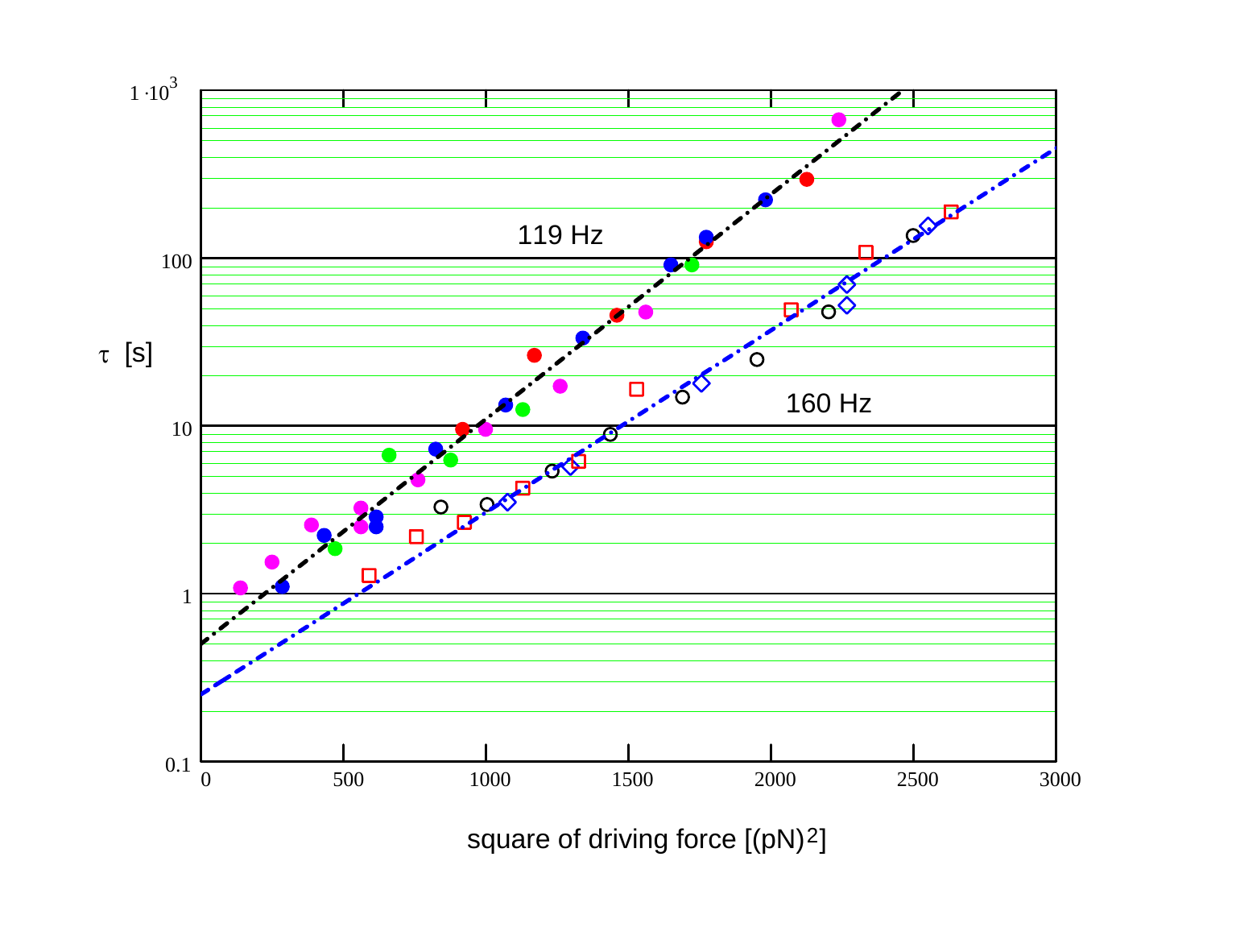}
\end{center}
\caption{ (From \cite{Schoepe2013}) Mean turbulent lifetimes as a function of the driving force at two different oscillator frequencies. Each data point is obtained from a time series, some of which lasted up to 36 hours. The straight lines are fits of Eq.\,\eqref{Eq:22} to the data. The data of the 119 Hz oscillator were taken at 4 different temperatures (in mK): red 403; blue 301; green 200; and violet 100. 
The data at 160 Hz are taken at 300 mK (black circles); at 30 mK with a mixture of 0.05\% $^3$He (red squares); at 30 mK with 0.5\% of $^3$He (blue diamonds). 
Note that the slopes $1/F_1^2$ and the intercepts $\tau _0$ are independent of temperature and $^3$He concentration, but both depend on the oscillation frequency. (Color figure online)}
\label{fig:15}       % Give a unique label
\end{figure}

The turbulent phases are found to be exponentially distributed as $\exp(-\,t /\tau)$, see Fig.\,\ref{fig:14}.
The mean lifetime $\tau $ increases very rapidly with increasing driving force $F$, namely as 
\begin{equation}
\tau (F) = \tau_0 \,\, \exp[\,(F/F_1)^2\,] \, , \label{Eq:21}
\end{equation}
see Fig.\,\ref{fig:15}. These results were confirmed by the Osaka group in experiments with a vibrating wire \cite{Osaka}.\\
From a fit of Eq.\,\eqref{Eq:21} to the data, we obtain for the 119 Hz oscillator $\tau _0$ = 0.5 s and $F_1$ = 18 pN, while for the 160 Hz oscillator we have $\tau _0$ = 0.25 s and $F_1$ = 20 pN. Although we have data only for two frequencies, it is possible that $F_1$ scales as $\sqrt {\omega }$ because $\sqrt {160/119}$ = 1.16 and 20/18 = 1.11. In the following, we will first concentrate on the properties of $F_1$ by using dimensional arguments. 

\begin{figure}
\begin{center}
\includegraphics[width=0.80\textwidth]{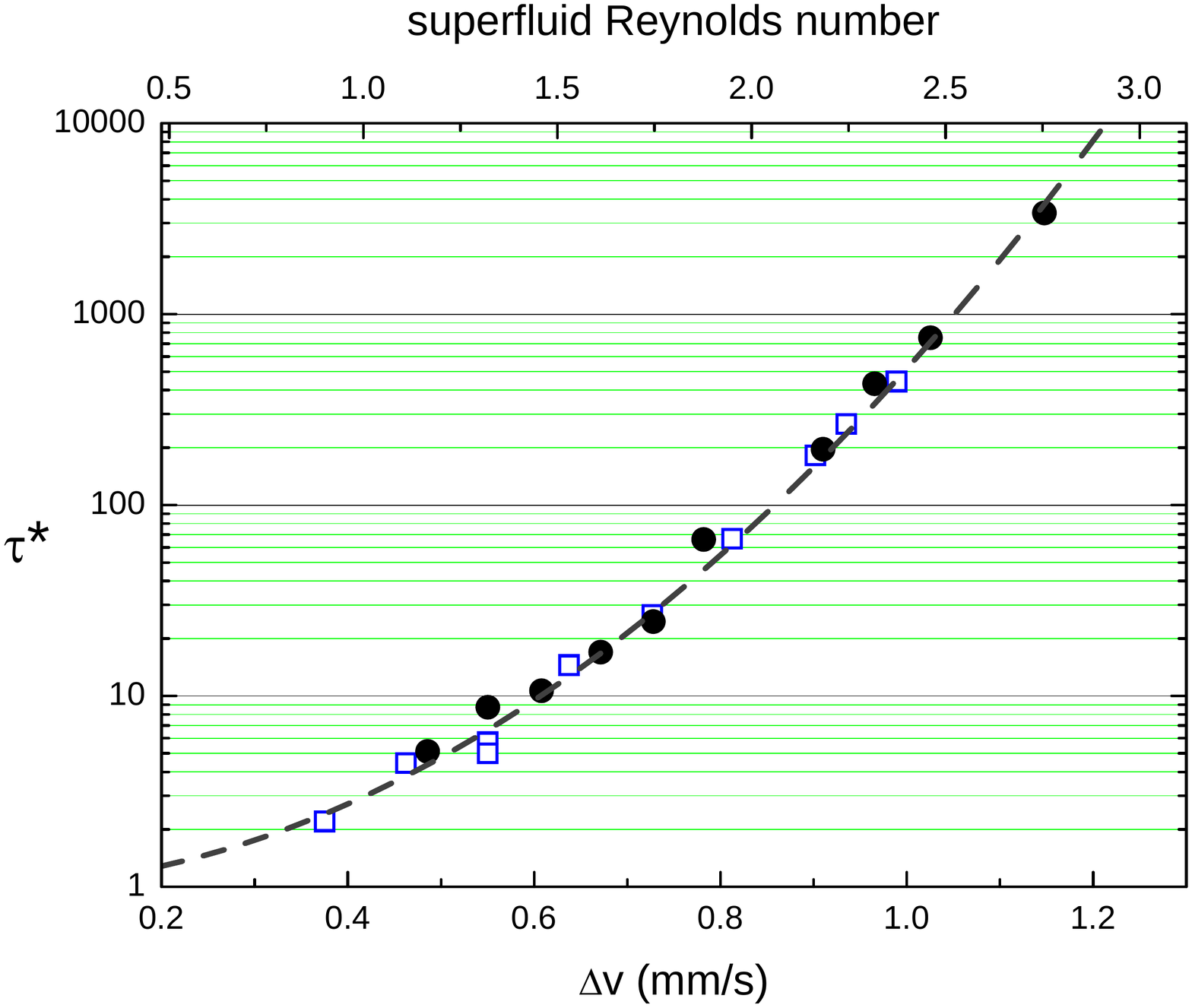}
\end{center}
\caption{(From \cite{Schoepe2015}) The normalized lifetimes $\tau^* = \tau/\tau_0$ as a function of $\Delta v = v - v_c$ for the 119 Hz oscillator at 301 mK (blue squares) and the 160 Hz oscillator at 30 mK  with 0.05\% $^3$He (black dots). Note the rapid increase of $\tau ^*$ by 3 orders of magnitude over the small velocity interval of ca. 0.7 mm/s. At the top axis, the corresponding values of the superfluid Reynolds number are given, see text. The dashed line is calculated from Eq.\,\eqref{Eq:27}. (Color figure online)}
\label{fig:16}       % Give a unique label
\end{figure}

The force $F_1$ must also depend on the density of liquid helium $\rho = 145$ kg/m$^3$, a length scale which we take as the radius $R = 0.12 $ mm of the sphere, and the circulation quantum $\kappa \approx 10^{-7}$ m$^2$/s.
The result is
$F_1 \propto \rho \,\kappa ^{3/2} \,R \,\sqrt{\omega} = \rho \,\kappa \,R\, \sqrt{\kappa\, \omega}$.
The latter expression is more appropriate because $\sqrt{\kappa\, \omega}$\, determines the critical velocity, see above.
Of course, we must allow for a numerical factor $c$ which can be obtained only from the experiment:

\begin{equation}  
F_1  = c\, \rho \, \kappa \,R \,\sqrt{\kappa \,\omega} \, . \label{Eq:22}
\end{equation}
From our experimental results for $F_1$ we find $c \approx 1.3$. We propose the following model for $F_1$.
We assume that a vortex ring of the size $R$ is shed by the sphere during each half-period $T/2 = \pi /\omega $. The power lost by the sphere amounts to the energy $E(R)$ of the ring per half-period. For $E(R)$ we take the usual formula, namely 
\begin{equation} 
E(R) = 1/2\,\, \rho \kappa ^2 \, R [\,\ln( 8R/a_0) - 7/4] \, .\label{Eq:23}
\end{equation}
The term in the brackets is roughly 14.
The dissipated power is then given by
\begin{equation}
\frac{1}{2} F_1 v_c = E(R)\omega/\pi \, , \nonumber
\end{equation}
i.e.,
\begin{equation}
F_1 = (14/2.8 \,\pi )\, \rho\,\kappa \,R\, \sqrt{\kappa \,\omega} = 1.6 \, \rho\,\kappa \,R\, \sqrt{\kappa \, \omega} \, . \label{Eq:24}\\
\end{equation}
This result agrees with Eq.\,\eqref{Eq:22} and with our experiment: $F_1$ scales as $\sqrt{\omega }$ and is independent of temperature, the numerical factor of 1.6 is close to our experimental value of 1.3. We interpret the number $n = F/F_1$ as the average number of vortex rings emitted per half-period. Our data in Fig.\,\ref{fig:15} lie in the range $0.7 \le n < 3$. 
In the regime where we observe the intermittent switching $\Delta v/v_c \le 0.03$ we may approximate $F$, see Eq.\,\eqref{Eq:4}, by

\begin{equation}
F \approx \frac{8\gamma}{3 \pi}\ 2 \,v_c \,\Delta v \, . \label{Eq:25}
\end{equation}
We proceed by changing variables from forces to velocities, and write 
\begin{equation}
n = \frac{F}{F_1} = \frac{(8/3\pi)\,2 \gamma  v_c \, \Delta v}{1.3\,\,\rho \,\kappa \,R\, \sqrt{\kappa \,\omega}} = \frac{\Delta v}{v_1} \, , \label{Eq:26}
\end{equation}
where the ``characteristic" velocity is $v_1 = 0.48 \,\,\kappa / R$ = 0.40 mm/s. We note that $\kappa /R$ is of the same order of magnitude as the self-induced velocity of a vortex ring of size $R$.
Using Eq.\,\eqref{Eq:26} we can plot the normalized lifetimes
\begin{equation}  
\tau^* (\Delta v) = \tau / \tau _0 = \: \exp \,(n^2) =  \: \exp[(\Delta v/v_1)^2] \, , \label{Eq:27}
\end{equation}
see Fig.\,\ref{fig:16}. From Eq.\,\eqref{Eq:25} it also follows that the force can now be written as a function of $n$: $F(n) = (8/3 \pi )\,2\, n \,\gamma \,v_1\,v_c$. 

The exponential dependence of $\tau $ on $(\Delta v / v_1)^2$ may be described by Rice's formula for Gaussian fluctuations crossing a given level \cite{Rice,Tik}. It is applied here to fluctuations of $\Delta v$ crossing $v_c$, where turbulence breaks down. The average number $N$ per unit time of crossings of a given level $C$ below a mean at zero with negative slope is proportional to $\exp[- C^2/2\sigma ^2]$, with $\sigma ^2$ being the variance of the fluctuations. We assume, that the breakdown occurs at the first crossing. Thus, we have $\tau \propto 1/N$. Comparing this with Eq.\,\eqref{Eq:27}, we set $C \equiv \Delta v$ and $ \sigma  \equiv v_1 / \sqrt{2}$ = 0.34 $\kappa /R$ = 0.27 mm/s. Therefore, $v_1$ determines the width of the Gaussian fluctuations of the velocity. These fluctuations cannot be attributed to the sphere, because at $\sim $ 0.1 K the relaxation time of the oscillator is larger than 1000 s. We thus must conclude, that the fluctuations are due to the flow velocity of the superfluid around the sphere.\\
The prefactor $\tau _0 $ is interpreted to be twice the average time between two crossings of the level $v_c$, i.e., for $\tau _0$ = 0.5 s or 0.25 s there are 4 or 8 crossings per second which means that the velocity of superfluid around the oscillating sphere fluctuates at an average frequency of 2 Hz or 4 Hz, respectively. This could not be detected in our experiments, nor can we firmly establish $\tau_0 (\omega )$ from those 2 values.\\

\begin{bf}
\subsection{Vortex shedding from the oscillating sphere and from a laser beam moving through a BEC}
\end{bf}

In this Subsection we compare
our own experiments in superfluid helium \cite{Schoepe2016} with those in a BEC as observed by other authors, where a moving laser beam sheds vortices above a critical velocity. In particular, the frequency $f_v$, with which vortices are shed, is found to be similar. Beginning with our experiments, we obtain from the average number $n$ of vortex rings shed per half-period, see Eq.\,\eqref{Eq:26},
\begin{equation}
f_v =  2 n f = \frac{2f \Delta{v}}{v_1} = a\,\,\Delta v \, , \label{Eq:28}
\end{equation}
where the coefficient $a$ = 2$f /v_1$. At $f$ = 119 Hz we obtain $a$ = 0.60 $\mu$m$^{-1}$ and at 160 Hz $a$ = 0.80 $\mu$m$^{-1}$. From $1/a$ we have a characteristic length scale, which is given here by $v_1 / 2f$ = 0.48 $\kappa /2fR$. This can be interpreted as the distance a vortex ring travels during one-half period.

\begin{figure}[tb]
\centerline{\includegraphics[width=0.8\textwidth]{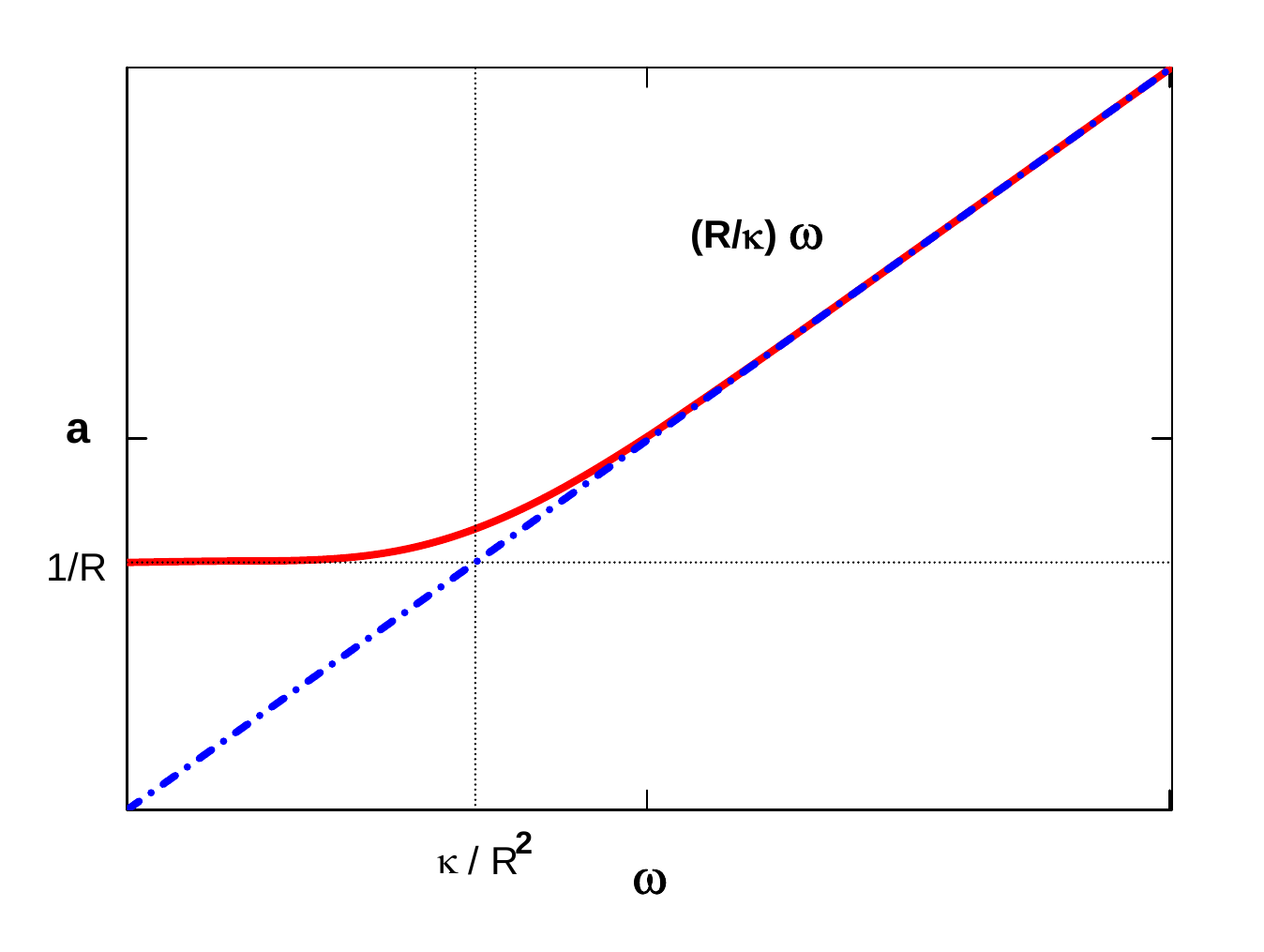}}
\caption{(From \cite{Schoepe2016}) Sketch of the coefficient $a$ of $f_v = a\, \Delta v $ as a function of the oscillation frequency $\omega$. At small frequencies the radius $R$ of the sphere is taken as the characteristic length scale, hence $a \sim 1/R$, whereas at large frequencies $a$ scales as $(R/\kappa)\,\omega$, see Eq.\,\eqref{Eq:28}. The characteristic frequency that marks the transition between both regimes is given by $\kappa /R^2$. Numerical factors of order 1 are neglected. (Color figure online)}
\label{fig:17} 
\end{figure}

The linear increase of $a$ with $f$ likely breaks down at very large frequencies, i.e., when the period becomes shorter than the time it takes a vortex ring to be shed. Then our picture of shedding of individual rings may no longer be applicable. Reconnections and annihilations near the surface of the sphere will become important. In the opposite limit, when $f$ goes to zero for uniform motion, it is clear that $f_v$ must remain finite because vortices can clearly be created also in the case of steady motion. Therefore, there must be a crossover from the linear frequency dependence to a constant value. In that case, the relevant length scale is assumed to be the radius $R$ of the sphere. We set $a \sim 1/R$ and obtain $a \sim$ 0.008$\mu$m$^{-1}$. This is a small but finite value instead of zero. For a sketch of the frequency dependence of $a$, see Fig.\,\ref{fig:17}.

Above we used analogous arguments to estimate the change of the critical velocity for oscillatory flow $v_c \sim \sqrt{\kappa \omega}$ when $\omega $ approaches zero for steady motion, see Section \ref{s.vc}. The characteristic length scale is assumed to change from the oscillation amplitude to the size of the sphere, reproducing the Feynman critical velocity $\sim \kappa /R$. The transition occurs at a characteristic oscillation frequency $\omega_k \sim \kappa/ R^2$. It is remarkable, that $\omega_k$ is the same (except for some numerical factor) as the frequency where the shedding $f_v$ has its transition, see above. Thus, our experiments with the sphere in $^4$He are clearly in the regime of oscillatory flow, concerning both $v_c$ and $f_v$.

Shedding of vortex dipoles in a stable and periodic manner has been observed recently at Seoul National University by moving a repulsive Gaussian laser beam steadily through a BEC of $^{23}$Na atoms \cite{Shin}. The shedding frequency $f_v$ is also given by $a \Delta v$, where now $v_c$ = 0.99 mm/s and $a$ = 0.25 $\mu$m$^{-1}$. Because the beam was moved steadily, we assume that the relevant length scale is given by the size of the beam 2$R$ = 9.1 $\mu$m, in accordance with the arguments presented above. Therefore, we estimate $a \sim 1/R$ = 0.22 $\mu$m$^{-1}$, in fair agreement with the experimental result, and $v_c \sim \kappa /R$ = 3.7 mm/s (where $\kappa$ = 1.7 $10^{-8}$ m$^2$/s for the $^{23}$Na BEC). If the beam would have been oscillating at a frequency substantially larger than $\omega_k \sim \kappa/R^2$ = 803 s$^{-1}$ (or 128 Hz), we would expect a shedding frequency $f_v$ proportional to $\omega$ and a critical velocity scaling as $\omega^{1/2}$, in accordance with our results in $^4$He.

Earlier experiments with a moving beam were performed by Ketterle's group \cite{Kett1,Kett2}. In that work, the beam was moved back and forth at frequencies below 200 Hz and the diameter of the beam was 10 $\mu$m. Although the motion was not a sinusoidal one, we can estimate $\omega_k  \sim $ 680 s$^{-1}$ (or 108 Hz), which is of the same order of magnitude as the applied frequencies, and therefore the frequencies are too close to the steady regime to show the frequency dependencies of both $f_v$ and $v_c$. More details of the work can be found in \cite{Kett3} where care was taken to distinguish between the critical velocity for heating the BEC due to phonon emission by the moving beam and that of vortex production. In our experiments in helium, the situation is simpler because of the very different ratio of the critical velocity to the speed of sound that is of the order of $10^{-4}$ in $^4$He, while in the BECs it is typically 0.1 to 0.5. Therefore, in our case the critical velocity is not affected by phonon emission.

There are several numerical investigations \cite{Jackson} confirming the linear increase of the energy dissipation above $v_c$ which is equivalent to a linear increase of $f_v$. That result agrees also with earlier work on the drag force at steady motion \cite{Frisch} from which also a linear behavior of $f_v$ can be inferred. Both the dissipation and the drag were found to be caused by vortex shedding above a critical velocity. This supports our general picture presented above. 

Recently, an oscillating object in a BEC has been investigated theoretically \cite{Makoto2}, but shedding frequencies have not been calculated so far. The frequency range in that work was limited to 0.2 $<\omega/\omega_k<$ 1.0, and therefore is also in the regime of steady flow, which is also evident from the frequency independence of the critical velocity for multiple vortex pair production $\approx \kappa /R $.

A very recent work by V.P. Singh et al. \cite{Singh} presents exciting new details of the effect of stirring a BEC with a laser beam.
If the stirrer is repulsive, the critical velocity is governed by vortex production, while for an attractive one, phonon emission is the relevant mechanism for heating. Both circular and linear motion of the beam were used. Vortex shedding frequencies were not an issue in that work but an estimate of the characteristic frequency $\omega_k$ gives a value of about 100 s$^{-1}$ which is of the same order of the rotation frequency, decomposed into 2 linear oscillations. Therefore, the flow regime was probably near the steady limit. 
More investigations on vortex shedding in BECs and in superfluid helium could demonstrate the similarity of both types of superfluids.\\

\begin{bf}
\subsection{\bf A new example of ``supertransient'' chaos}
\end{bf}

In a numerical investigation of the 2-dimensional Gross-Pitaevskii equation \cite{Reeves} a new superfluid Reynolds number has been introduced to describe the onset of vortex shedding:
\begin{equation}
Re_s \equiv \frac{\Delta v\, D}{\kappa} \, . \label{Eq:29}
\end{equation}
%By replacing $\Delta v$ by $Re_s$
Solving for $\Delta v$ 
and setting the characteristic length scale $D = 2R$, we obtain from Eq.\,\eqref{Eq:27}
\begin{equation}
\tau^* = \,\exp\,[(c\, Re_s)^2] \, , \label{Eq:30}
\end{equation}
where c = 1.04. The accuracy of c is determined by the accuracy of several numerical factors in Eq.\,\eqref{Eq:26} and is estimated to be about 10\%. That means within our accuracy we may as well set c = 1.

In Fig.\,\ref{fig:16} the dependence of $\tau^*$ on $Re_s$ is shown at the top axis (note that $Re_s$ = 0 corresponds to $\Delta v$ = 0 and $\tau^*$ = 1). 
All values of $Re_s$ are below 3. 
An extension towards larger values of $Re_s$  and $\tau^*$ would require time series lasting much longer than our longest one of 36 hours. 
Therefore, the validity of Eq.\,\eqref{Eq:30} cannot safely be extrapolated. 
Nevertheless, we can state that the fast increase of $\tau^*$ with $Re_s$ is a new example of supertransient chaos, where the lifetime of the turbulent state grows faster than exponentially with the Reynolds number \cite{Tel}, here for the first time in a superfluid.

Finally, it should be mentioned that an interesting conclusion can be drawn from a comparison of Eq.\,\eqref{Eq:27} with Eq.\,\eqref{Eq:30}, from which we find
\begin{equation}
Re_s = \,n. 
\end{equation}
That means, in our experiment the superfluid Reynolds number is given simply by the number of vortex rings that are shed from the sphere in one-half period of oscillation. This is a surprisingly simple result. In a much different context an equally simple result for $Re_s$ has been obtained theoretically in 2D superfluid turbulence \cite{Ash}.
\medskip

\section*{SUMMARY}
Summarizing our experiments with the oscillating sphere in superfluid $^4$He we conclude, that we have a fairly good understanding of the flow in the various regimes. 
However, several of our results require more work, both experimental and theoretical. 
In particular, a theory of the transition to turbulence for oscillatory flow is needed in order to understand the origin of the critical velocity $v_c(\omega) $ and to calculate the vortex density $L$, especially at $v_c$. 
Furthermore, the frequency dependence of the rms of the Rayleigh distribution of the laminar phases $v_w (\omega )$, and the lifetimes of the turbulent phases $\tau _0 (\omega )$ need to be established and explained. 
Still surprising and unexplained, even after 15 years of publication, is the linear temperature dependence of the laminar drag in the dilute $^3$He - $^4$He mixtures. 
On the experimental side, it would be interesting to use other spheres, having a different size and a smoother surface. 
An extension of our method to superfluid $^3$He will probably yield  more surprises. 
Finally, more experiments on vortex shedding in BECs should test our predictions concerning the diameter of the beam and its oscillation frequency.

\begin{acknowledgements}
We acknowledge early contributions by K. Gloos, J. Simola, and J. Tuoriniemi (then all at Helsinki University of Technology, Finland), who invented the technique to oscillate a levitating particle, and shared their expertise with us. 
Experiments with a levitating sphere immersed in superfluid helium were conducted at Regensburg University from 1993 on by H. Barowski and J. J\"ager. H. Kerscher was a co-worker on the experiment in 2000. 
M. Bleher (Federal Office for Radiation Protection (Bf\,S), Braunschweig) performed the measurements of the dose rate of the natural background radiation in our laboratory. 
A helpful discussion with the late Shaun Fisher is gratefully acknowledged. 
Support and encouragement came from M. Krusius (Aalto University, Helsinki) and from K.F. Renk (Regensburg University). 
R.H. was supported by the Academy of Finland, and M.N. by the Deutsche Forschungsgemeinschaft. 
W.S. is grateful to the O.V. Lounasmaa Laboratory (Aalto University) for the warm hospitality experienced during many visits.
\end{acknowledgements}

\section*{APPENDIX A}
\begin{figure}[!b!]
\centerline{\includegraphics[width=0.6\textwidth]{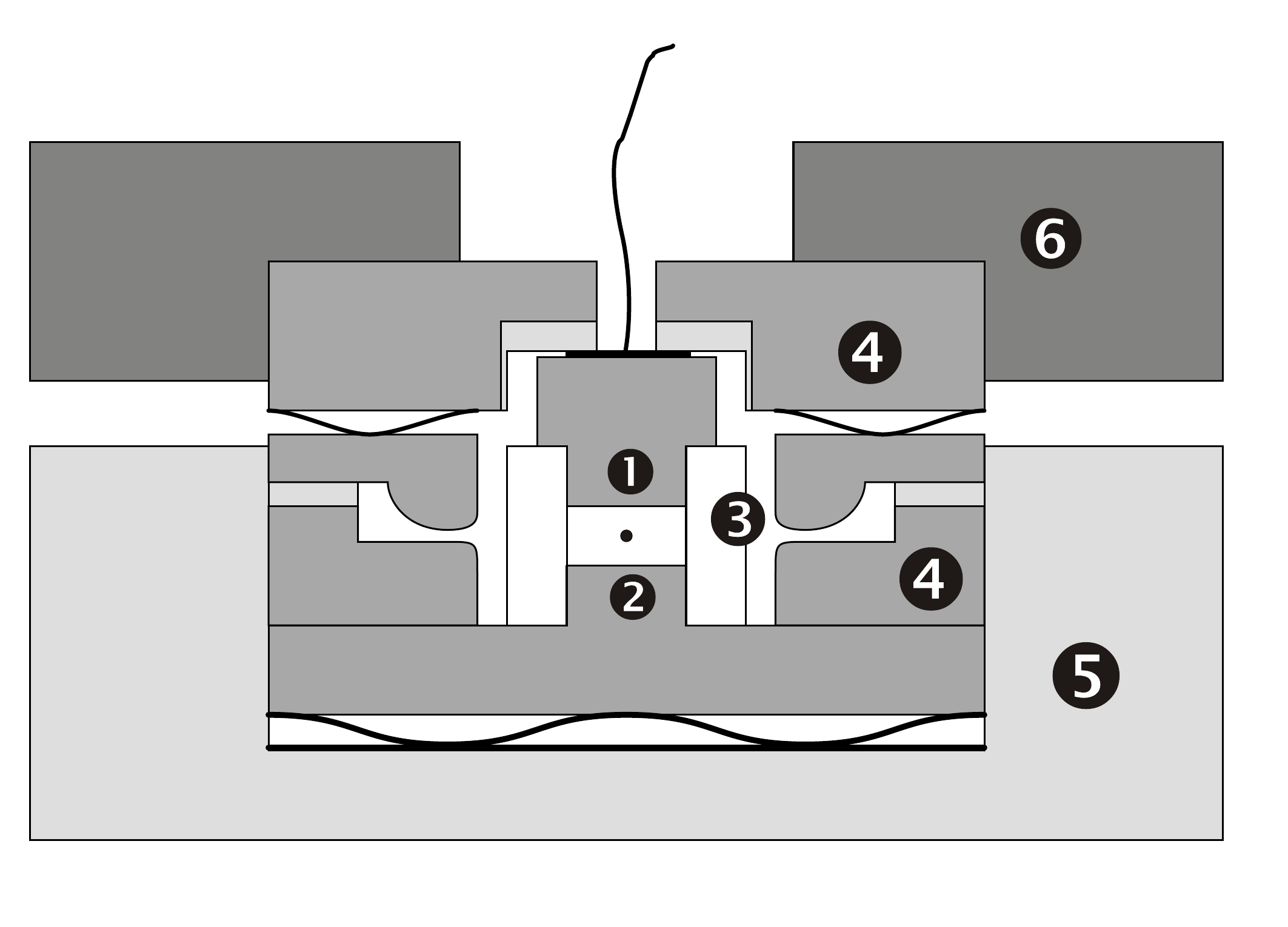}}
\caption{(From \cite{NiemetzSchoepe2004}) Schematic of the measuring cell. Niobium capacitor with 1 mm gap and 2 mm diameter (1, 2); quartz spacer (3); Nb shielding (4); plastic (5); brass (6).}
\label{fig:18} 
\end{figure}

\subsection*{\textbf{Experimental details}}

\begin{figure}[t]
\centerline{\includegraphics[width=0.4\textwidth]{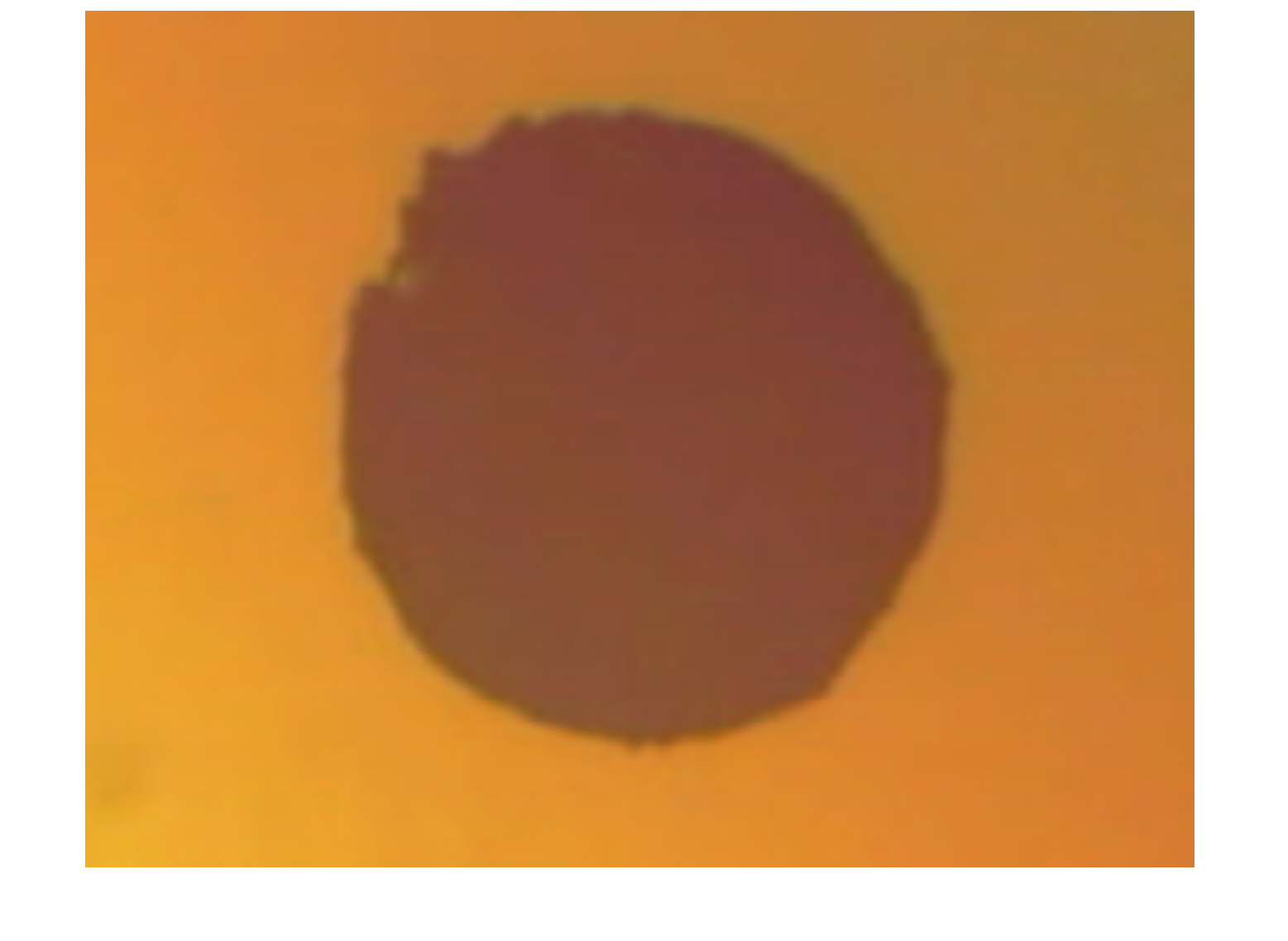}}
\caption{(From \cite{NiemetzSchoepe2004}) The sphere is made of ferromagnetic $\mathrm{SmCo_5}$ embedded in a polymer matrix. While the overall structure is rather spherical, surface protrusions are visible. (Color figure online)}
\label{fig:19} 
\end{figure}

\begin{figure}[h]
\centerline{\includegraphics[width=0.8\textwidth]{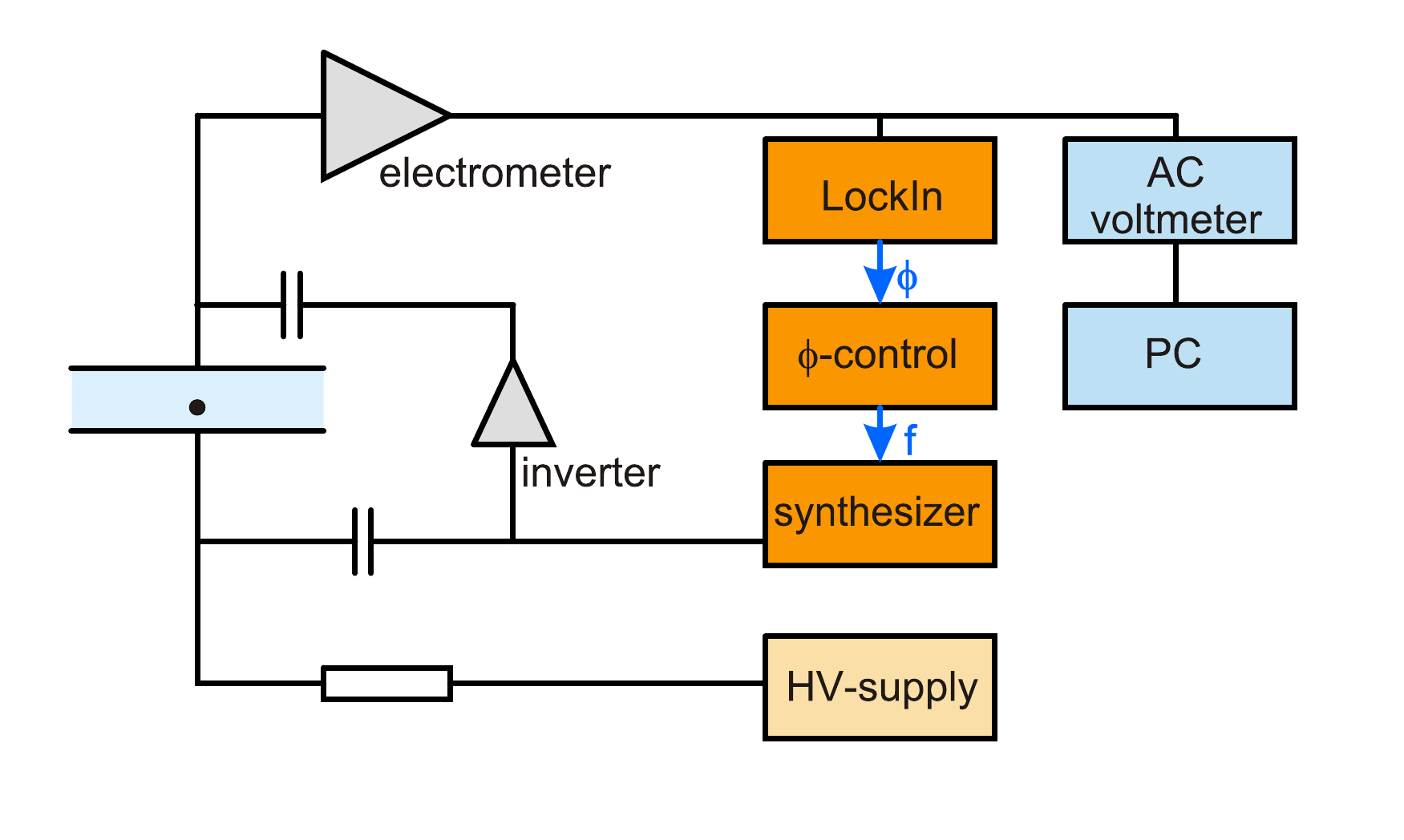}}
\caption{(From \cite{NiemetzSchoepe2004}) The oscillations of the sphere are detected by the electrometer. 
The phase is locked at resonance by adjusting the frequency of the synthesizer. 
The inverter is used to null the capacitively coupled pick-up. 
The high-voltage supply is disconnected when the sphere is levitating. (Color figure online)}
\label{fig:20} 
\end{figure}

1. The experimental setup:
The measuring cell is shown in Fig.\,\ref{fig:18}. The central part is the parallel plate capacitor made of niobium with the sphere in between the electrodes.
The sphere of diameter 240 $\mu $m ($\pm $ 8 $\mu $m) as seen through a microscope is depicted in Fig.\,\ref{fig:19}.  
A schematic of the electronics is shown in Fig.\,\ref{fig:20}.

\medskip
\noindent
2. The experimental procedure: The evacuated measuring cell is cooled down. 
Just before the niobium capacitor becomes superconducting we apply several 100 volt to the capacitor. 
The surface charge of the sphere puts it into an up-and-down motion that is detected by the electrometer. 
The motion stops when the capacitor plates are superconducting. 
The high-voltage supply is disconnected.\\ 
After some waiting time at 4 K (usually overnight) we detect the resonance frequency simply by knocking gently at the cryostat that is isolated from vibrations of the laboratory. 
The signal from the vibrating sphere is inspected on a scope to make sure that there are no beats, and that the signal is sinusoidal within our range of amplitudes. 
The oscillator has a nonlinear return force (Duffing oscillator) with an amplitude dependent resonance frequency \cite{JaegerSchudererS1995a}. 
We continue to cool the cell further down.\\ 
Next, we determine the background damping of the sphere in the empty cell from the exponential free decay. 
Finally, we fill the cell with superfluid helium and begin the data acquisition by applying small driving forces at constant temperature to measure the linear damping. 
A free exponential decay of the signal gives the time constant, as described in the Introduction. 
Every morning we check for stability of the charge by reproducing some data of the day before. 
Usually the charge is quite stable.\\ 
A change of the resonance frequency requires a new levitation status that is achieved first by evacuating the cell while heating it above $T_c$ of Nb and then following the same procedure as above.

\section*{APPENDIX B}

\subsection*{\textbf{The linear drag force due to $^3$He impurities in dilute mixtures}}

\begin{figure}[h]
\centerline{\includegraphics[width=0.6\textwidth]{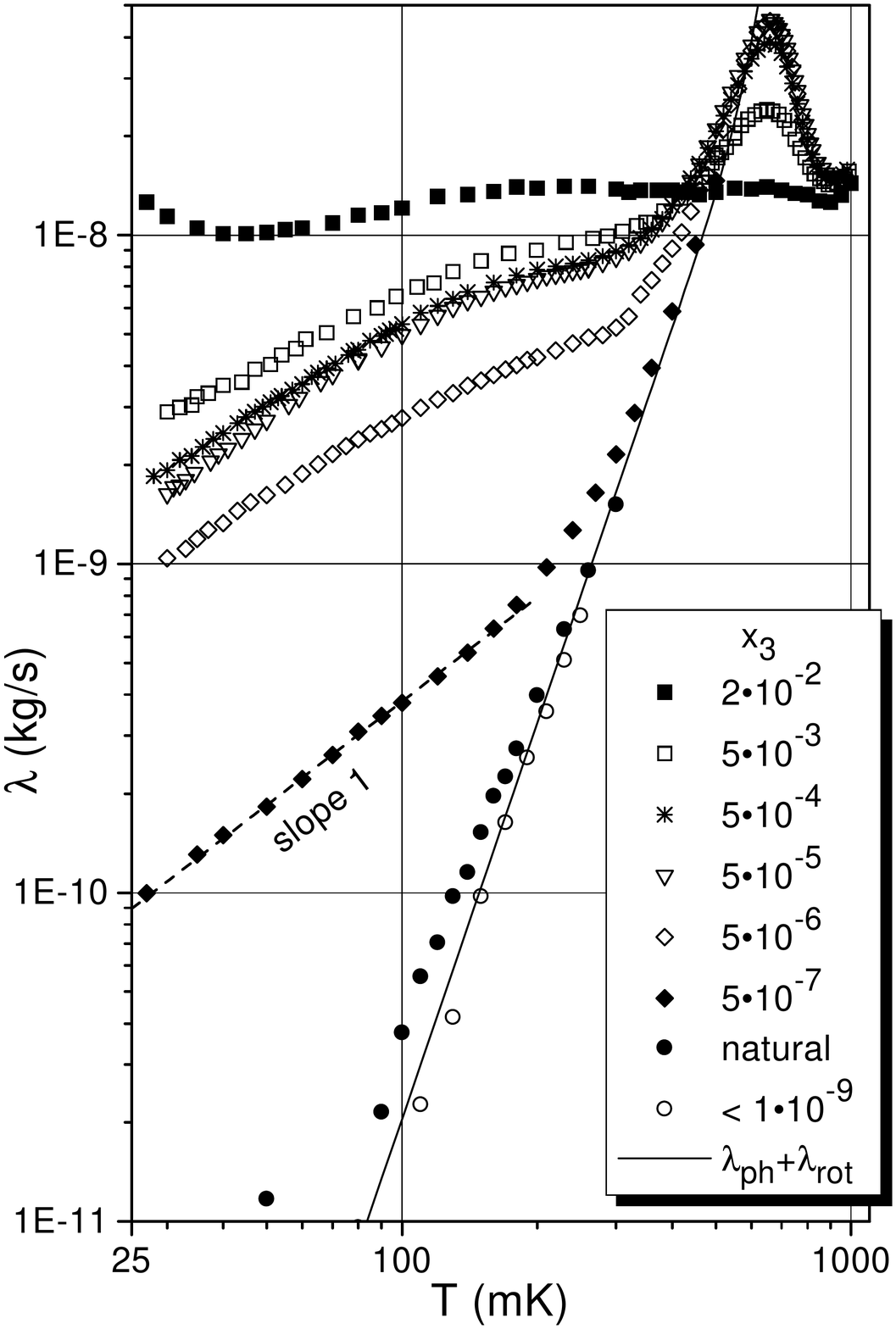}}
\caption{(From \cite{NiemetzKerscherSchoepe2001a}) Temperature dependence of the linear drag coefficient $\lambda $ for various $^3$He concentrations x$_3$, for ``natural'' and for purified $^4$He. 
Note that from 5$\cdot $10$^{-6}$ to 5$\cdot $10$^{-7}$ $\lambda $ scales as $x_3 \cdot T$. 
The solid line is calculated for ballistic quasiparticle scattering in pure $^4$He. (Background damping subtracted).}
\label{fig:21} 
\end{figure}

We have extended our experiments to dilute $^3$He - $^4$He mixtures in order to measure the drag caused by the $^3$He impurities. 
Especially in the limit of ballistic scattering at low concentrations and low temperatures, we had expected the drag due to $^3$He - scattering to be given simply by 
\begin{equation}
\lambda  = c\,\rho \langle v\rangle \pi R^2\, ,
\end{equation}
where $c$ is a numerical factor depending on details of the scattering mechanism, $\rho $ is the density of the $^3$He atoms having an effective mass $m_3^* = 2.4\, m_3$, and a thermal velocity   
$\langle v \rangle = (3 kT/m_3^*)^{1/2}$. 
This would lead to $\lambda \propto x_3 \cdot T^{1/2}$, where $x_3$ is the $^3$He concentration. 
From Fig.\,\ref{fig:21} we see that instead $\lambda $ scales as $x_3 \cdot T$. 
So far, there is no explanation for this peculiar result that we know of \cite{BBP}.
For more details of our work we refer to the original publication \cite{NiemetzKerscherSchoepe2001a}. 

%\nocite{*}

%\bibliography{OscillatingSphere}   % name your BibTeX data base


% etc
%\end{thebibliography}

\end{document}